\documentclass[final,5p,times,twocolumn]{elsarticle}

\usepackage{lineno}
\usepackage{graphicx}
\usepackage{amssymb}
\usepackage{pdflscape}
\usepackage[english]{babel}
\usepackage[dvipsnames,svgnames,x11names]{xcolor}
\usepackage{booktabs,tabularx}
\usepackage{multirow}
\usepackage{subfig} 
\usepackage{hyperref}
\usepackage{amsmath}
\usepackage{commath}
\usepackage[english]{babel}
\usepackage[ruled]{algorithm2e}
\SetEndCharOfAlgoLine{}
\usepackage[leftmargin=6em,rightmargin=12em,indentfirst=false]{quoting}
\usepackage{siunitx}
\usepackage{comment} 

\usepackage[final]{changes}
\usepackage{float}

\usepackage{lineno}
\usepackage{tikz}
\usepackage{comment} 
\usepackage[export]{adjustbox}

\usepackage{graphicx}
\usepackage[utf8]{inputenc}
\usepackage[export]{adjustbox}
\usepackage{wrapfig}
\usepackage{dcolumn}

\makeatletter
\newcolumntype{T}[3]{>{\textfont0=\the@{#1}{#2}{#3}}c<{\DC@end}}
\makeatother

\usepackage{pgfplots}
\pgfplotsset{width=10cm,compat=1.9}

\usepackage{array}

\newcolumntype{L}[1]{>{\raggedright\let\newline\\\arraybackslash\hspace{0pt}}m{#1}}
\newcolumntype{C}[1]{>{\centering\let\newline\\\arraybackslash\hspace{0pt}}m{#1}}
\newcolumntype{R}[1]{>{\raggedleft\let\newline\\\arraybackslash\hspace{0pt}}m{#1}}

\usepackage{subfiles}

\usepackage{todonotes}

\journal{Buildings}
\begin{document}
	
\begin{frontmatter}

\title{Humans-as-a-sensor for buildings: Intensive longitudinal indoor comfort models}

\author{Prageeth Jayathissa, Matias Quintana, Mahmoud Abdelrahman, and Clayton Miller\,$^{*}$}

\address{Building and Urban Data Science (BUDS) Lab, National University of Singapore (NUS), Singapore}

\address{$^*$Corresponding Author: clayton@nus.edus.sg, +65 81602452}

\begin{abstract}
Evaluating and optimising human comfort within the built environment is challenging due to the large number of physiological, psychological and environmental variables that affect occupant comfort preference. Human perception could be helpful to capture these disparate phenomena and interpreting their impact; the challenge is collecting spatially and temporally diverse subjective feedback in a scalable way. This paper presents a methodology to collect intensive longitudinal subjective feedback of comfort-based preference using micro ecological momentary assessments on a smartwatch platform. An experiment with 30 occupants over two weeks produced 4,378 field-based surveys for thermal, noise, and acoustic preference. The occupants and the spaces in which they left feedback were then clustered according to these preference tendencies. These groups were used to create different feature sets with combinations of environmental and physiological variables, for use in a multi-class classification task. These classification models were trained on a feature set that was developed from time-series attributes, environmental and near-body sensors, heart rate, and the historical preferences of both the individual and the comfort group assigned. The most accurate model had multi-class classification F1 micro scores of 64\%, 80\% and 86\% for thermal, light, and noise preference, respectively. The discussion outlines how these models can enhance comfort preference prediction when supplementing data from installed sensors. The approach presented prompts reflection on how the building analysis community evaluates, controls, and designs indoor environments through balancing the measurement of variables with \emph{strategically asking for occupant preferences} in an intensive longitudinal way. 
\end{abstract}

\end{frontmatter}


\section{Introduction}
Many office workers are familiar with the battle of the thermostat, or that co-worker who talks loudly on the phone. Many researchers in indoor comfort are also aware of the high rates of discomfort amongst office workers \citep{frontczak2012quantitative,sakellaris2016perceived}. Vast global efforts have been undertaken to evaluate this discomfort, and with that knowledge, build models that can be used for the design and control of buildings. In the realm of thermal comfort, for example, two dominant models are in use. The first is the \emph{Predicted Mean Vote (PMV)} that models comfort based on heat transfer characteristics between the human and their surrounding environment \citep{fanger1970thermal}. The other, more modern version, is the \emph{Adaptive Comfort} model that includes the human adaptability to climate, drawing a linear relationship between the indoor and outdoor environments \citep{de1998developing}. 

\begin{figure*}
\begin{center}
\includegraphics[width=\textwidth, trim= 0cm 0cm 0cm 0cm,clip]{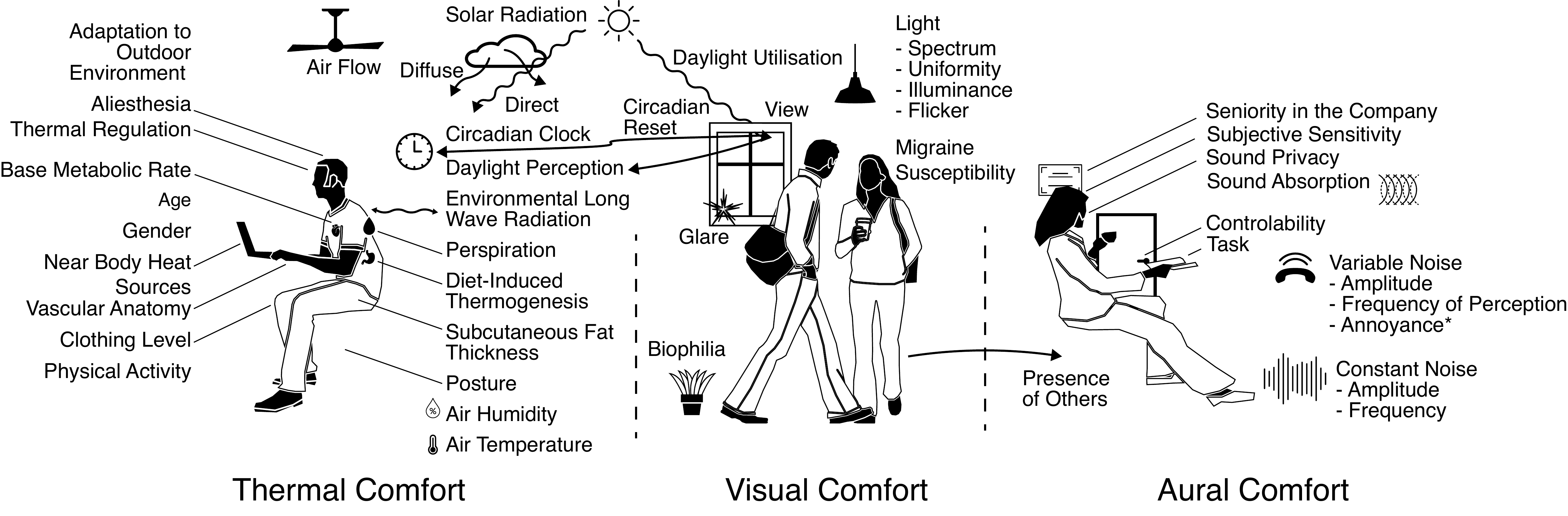}
\caption{Graphical review of physiological, psychological and environmental factors influencing human comfort. 
Thermal - clockwise from top left: adaptation to outdoor environment \citep{de1998developing}, air flow \citep{schiavon2017thermal}, solar radiation \citep{hodder2007effects}, circadian rhythm \citep{krauchi2002circadian},  daylight perception \citep{chinazzo2019daylight}, environmental long wave radiation \citep{halawa2014impacts}, perspiration \citep{fukazawa2009differences}, diet-induced thermo-genesis \citep{swaminathan1985thermic}, subcutaneous fat thickness \citep{kingma2012thermoneutral}, posture \citep{tikuisis1996effect}, temperature/humidity \citep{fanger1970thermal} \citep{gagge1967comfort}, physical activity \citep{gaesser1975muscular}, clothing level \citep{havenith2015database}, vascular anatomy \citep{johnson2010thermoregulatory}, near body heat sources \citep{zhang2014laptop}, gender \citep{karjalainen2012thermal}, age \citep{huang2010influence}, basal metabolic rate \citep{havenith2002personal}, thermal regulation \citep{frank1999relative} \citep{boulant2000role}, alliesthesia \citep{cabanac1971physiological}. \hspace{\textwidth}
Visual - clockwise from top left: circadian calibration \citep{brainard2001action}, daylight \citep{perez1990modeling}, view \citep{leather1998windows}, spectrum \citep{dai2016spectral}, uniformity  \citep{slater1990illuminance}, illuminance  \citep{baron1992effects}, flicker \citep{wilkins1989fluorescent}, susceptibility to migraines \citep{main1997photophobia}, biophilia \citep{yin2018physiological}, glare \citep{hopkinson1972glare}. \hspace{\textwidth}
Aural - clockwise from top: seniority in a company \citep{pierrette2015noise}, subjective sensitivity \citep{job1988community}, sound privacy \citep{kim2013workspace}, sound absorption \citep{templeton2014acoustic}, controllability \citep{lee2010can}, task \citep{kjellberg1991noise}, variable and constant noise \citep{banbury2005office}.}
\label{fig:variables}
\end{center}
\end{figure*}

The underlying issue with modelling human comfort is the sheer number of variables present and the difficulty in accurately measuring them. Figure \ref{fig:variables} highlights this issue by detailing a list of studied physiological, psychological, and environmental variables that influence thermal, visual, and aural comfort. While the empirical models in the academic literature are capable of incorporating a handful of these variables, the exclusion of the rest can cause significant errors. One reason is that the interrelationship between different indoor environmental parameters is not well-known \citep{Asadi2016}. \emph{It was shown in a recent study that the lowest indoor environmental satisfaction factor drives the overall satisfaction \citep{Tang2020}}. For example, while one can measure the temperature and humidity of a room; the type of meal a person ate, and even the spices present in the meal, can put the human body in a different state of thermal perception \citep{henry1986effect,swaminathan1985thermic}. Furthermore, most of the studies that depend on measuring environmental variables using mobile carts with mounted sensors \citep{Webster2007,Choi2012,Kim2013} or low-cost continuous sensing sensors \citep{Parkinson2019} face problems related to the accuracy and calibration \citep{Nicol2011}. While being comprehensive in capturing most comfort-related factors, Figure \ref{fig:variables} excludes literature about physical and mental ailments, which further adds variance to the models.

It is, therefore, not surprising to find that preference prediction models with only a handful of the factors in Figure \ref{fig:variables} have low accuracy. For example, the previously mentioned PMV model uses personal and environmental parameters such as temperature, humidity, mean radiant temperature, air movement, and clothing and metabolism levels to predict thermal comfort.  A recent analysis showed that this model is \emph{only accurate 34\% of the time} \citep{cheung2019analysis}. In the control of real buildings, these models are further simplified, and it is usually the only variables of temperature, illuminance, and noise levels that are used to evaluate thermal, visual, and aural comfort, respectively. \\
 
\subsection{Can longitudinal human perception feedback supplement sensors?}
The human nervous system detects sensation and converts it into thoughts and feelings, which are the very foundation of the word \emph{comfort}. What if occupants in buildings were asked about their subjective preference in spaces, instead of only measuring environmental variables and using them to infer comfort? Collecting enough comfort preference feedback from a single person over days or weeks would take advantage of a human's ability to evaluate dozens of variables simultaneously, including those that are difficult to measure. How can this type of methodology be accomplished in a scalable way without annoying occupants too much or inducing survey fatigue? Can this approach provide insight into comfort problem areas that contemporary sensors are too expensive or problematic in implementation? 

The goal of this paper was to test the ability of an intensive longitudinal method to capture numerous environmental feedback data from experimental participants in a field setting. This study uses \emph{Micro Ecological Momentary Assessments (EMA)} as a subjective feedback methodology that overcomes many of the challenges presented by traditional methods \citep{stone2007historical}. Micro-EMA is a method of using a smartwatch interface to prompt and collect momentary, \emph{right-here-right-now} subjective feedback from a single person over several weeks \citep{intille2016muema}. Receiving a large amount of feedback from a single person in a diversity of spaces and comfort exposures provided the ability to understand the comfort \emph{preference tendencies} of a person. It is proposed that these behavioural tendencies can be used to segment people into groups related to how they perceive their environment. Grouping people with similar comfort preferences could, therefore, increase the accuracy of predicting where a person will be comfortable and what the system can to respond without additional sensors. Additionally, collecting large amounts of subjective preference data from numerous people in a particular space can characterise the comfort-related attributes of that space \emph{to supplement data being collected from the sensors installed}. If technically scalable and not too disruptive to an occupant, using \emph{humans-as-a-sensor} in buildings could change the way post-occupancy evaluations, building and system design, and controls and automation are done. There would be opportunities for people to provide feedback for short-term episodic uses (days or weeks), for building commissioning or long-term (months or years), and for continuous system control and management. This work complements the momentum from other disciplines focused on the use of humans as sensors for applications in detection of events using social media data \citep{Wang2014-qj}, for detecting emergencies \citep{Avvenuti2016-qy}, and for cybersecurity \citep{Vielberth2019-nl}. \\

\begin{figure*}
\begin{center}
\includegraphics[width=\textwidth, trim= 0cm 0cm 0cm 0cm,clip]{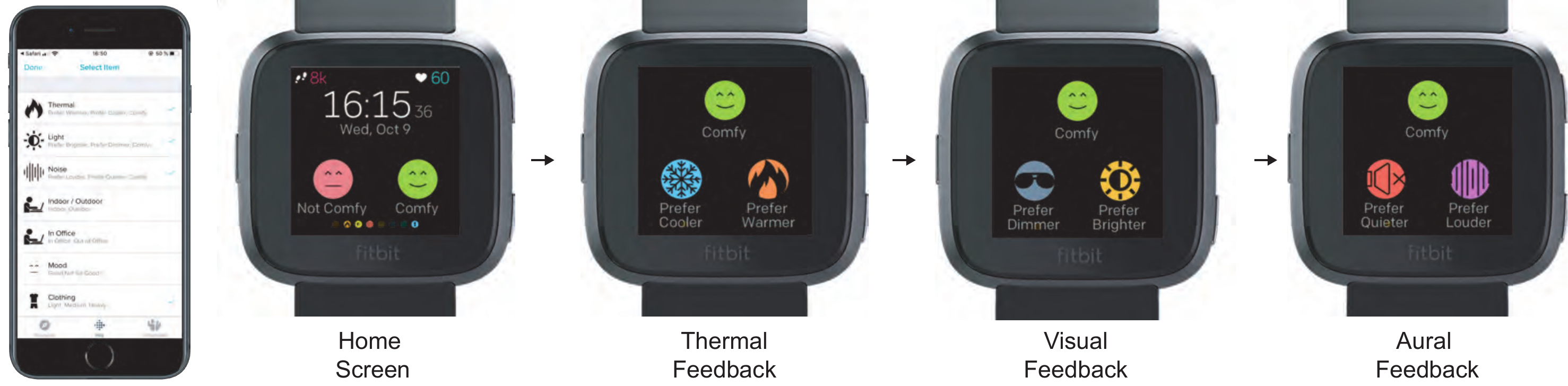}
\caption{
The cozie watch-face, built on the Fitbit smartwatch platform was used to collect subjective feedback. The phone that is paired with the Fitbit can be used to set up additional questions. 
}
\label{fig:cozie}
\end{center}
\end{figure*}

\subsection{Paper overview}
This paper presents how high-frequency micro-EMA, combined with sensor data time-series analysis, can enable the evaluation, control, and rethinking of the design of indoor environments. Section \ref{ch:literature} first gives a more detailed overview of foundational work in indoor preference capture and modelling and the novelty being proposed. Section \ref{ch:method} provides a comprehensive explanation of the design and deployment of a smart watch-based subjective preference data collection and environmental variable measurement system. Section \ref{ch:results} details the results from a field-based implementation at [removed for anonymization] and the testing of various preference models based on intensive longitudinal data. Finally, Section \ref{ch:discussion} and \ref{ch:conclusion} discusses integration methods in buildings, limitations, future work, and details on how to reproduce the study using open data and code.

\section{Background and novelty}
\label{ch:literature}
This work builds upon previous literature focused on the measurement of factors that may influence thermal, visual, and aural comfort in the built environment. These modelling techniques are converged with an intensive longitudinal experience sampling technique that is common in the medical and psychological communities, but only emerging in the analysis of buildings. This section covers previous work in the building context using intensive longitudinal data and an overview of the novelty of the work in this paper as compared to the literature.\\

\subsection{Indoor environmental comfort variables and models}
There are generally two models types used in the literature for indoor comfort assessment: 1) objective-subjective, and 2) objective-criteria \citep{Heinzerling2013}. Which method to use is decided based on the aim of the evaluation. On the one hand, the \emph{objective-subjective} model combines the indoor environmental measurements from sensors with the subjective feedback from users, mostly in the form of post-occupancy evaluation (POE) surveys \citep{Ncube2012,Wong2008,Lai2009,Cohen2001}. On the other hand, the \emph{objective-criteria} model is used in ranking or rating a building by comparing the indoor measurements from IEQ sensors with building performance measurement protocols such as LEED or WELL certifications \citep{Kim2013}. Both of the methods have drawbacks both in measuring the environmental data as well as surveying occupants \citep{Heinzerling2013}.



In terms of environmental measurements, work has been done that used accurate sensors that were mounted on movable carts \citep{webster2007ufad,Kim2013}. However, these sensors were not affordable to all building operations scenarios \citep{Heinzerling2013}. The affordability challenge was met using low-cost continuous sensing sensors that required frequent calibration \citep{Parkinson2019}. Nevertheless, the location of these sensors in buildings, and interpolation of the readings still represented a challenge in the literature, given the fact that indoor spaces are heterogeneous \citep{jin2018automated}. On the other side of the spectrum, surveys pose some problems related to questions, e.g. what to ask, whom to ask, and how to interpret the results \citep{Heinzerling2013}. Additionally, Porter et al. \citep{porter2004multiple} discussed the term \emph{survey fatigue} in which users feel overwhelmed by questions that may lead to a misrepresentation in responses and reduced response rates. 


A related area of recent focus is the use of wearable and infrared radiation sensors to capture \emph{near-body} physiological data that define the environmental conditions close to or at the skin surface of an occupant. A recent study focused on creating personalised comfort models from these data in the context of field-based deployment on 14 subjects \citep{Liu2019-pi}. This deployment and the models produced used wrist and ankle skin temperature from several sensors placed on the participants and a smartphone application to collect surveys. Further work in the indoor context showed that both wearable sensors and infrared radiation cameras led to a 3-4\% increase in accuracy of thermal comfort sensation prediction, marginally justifying the cost of implementation in a field setting \citep{Aryal2019-bx}.\\

\subsection{Ecological momentary assessments (EMA)}
The next area of background focuses on the challenge of collecting large amounts of longitudinal data from a person. Many fields of study have relied upon the ecological momentary assessment \citep{stone2007historical} methodology to meet this challenge. This method is a type of intensive longitudinal experience sampling most often utilised in studying human behaviour. The word \emph{ecological} describes that fact that the measurement is taken in the subjects' natural environment without impacting their task at hand. The word \emph{momentary} pertains to the fact that feedback is requested at the moment of experience, as opposed to asking a subject to recall a past experience. And finally, the \emph{assessments} are not static one-off outcomes but occur over time, thus accounting for temporal dynamics. Traditional models found in literature such as surveys are insufficient as their sampling rates are low, require the occupant to completely stop their task at hand to focus on the survey, and in many cases, ask for a recollection of past experiences. There is the further issue of survey fatigue \citep{porter2004multiple} and even when willing to participate, there is a concern about how accurate their responses are \citep{Clear2018}. The use of a smartwatch for data collection, coined \emph{micro ecological momentary assessments}, has been shown to be so user-friendly that it does not significantly disrupt any ongoing activity \citep{intille2016muema}. Furthermore, an eight-fold increase in sampling frequency can be obtained, in comparison to smartphone use, without burdening the user. Recent work has used ecological momentary assessments to assess the built environment through the use of smartphones \citep{engelen2019understanding}. While such applications are a step in the right direction, they were only able to collect eight feedback points per occupant, which is insufficient for time-series analysis. \\

\subsection{Similar work in intensive longitudinal data collection in the built environment}
Intensive longitudinal methodologies have begun to emerge as a way to characterise occupants for various built environment objectives. In the urban context, several studies have deployed sensors on people to understand their experiences across their daily lives. A large study based in Singapore used thousands of wearable sensors in populations of students to discover travel patterns \citep{Monnot2016-fl}, collect information about thermal parameters \citep{Wilhelm2016-il}, and even infer the impact of public spaces on happiness \citep{Benita2019-dl}. Work has been done in a controlled outdoor field study to understand the impact of the urban context on various emotions and physiological responses of human \citep{Ojha2019-jx}. In the indoor setting, targeted work on collecting longitudinal data for more specific purposes has also emerged. The previously mentioned wearable study focusing on thermal comfort collected numerous data from the 14 participants over the 2-4 week study \citep{Liu2019-pi}. Another recent study that deployed a cyber-physical system to collect longitudinal data in offices focused on occupant concentration \citep{Rahaman2020-gt}. The work in this paper is most directly related to previous work in collecting longitudinal comfort feedback from smartphone interfaces for the allocation of activity-based workspaces \citep{Sood2020-vz} and through a sustainability tour in a university campus building \citep{Sood2019-af}.\\

\subsection{Novelty of proposed approach}
Despite the momentum in field-based intensive longitudinal methodologies, there are still several barriers to their implementation in real-world settings. Not the least is the challenge of getting human occupants to give data for comfort surveys, install applications, or wear devices. Working from this knowledge, the authors developed \emph{cozie}, as seen in Figure \ref{fig:cozie}, an open-source, smartwatch clock-face designed to conduct micro-EMA surveys for high-frequency data collection \citep{jayathissayour}. The application is open-sourced and free to download and use on the Fitbit gallery \footnote{\url{https://cozie.app/}}.

\begin{figure*}[h!]
\begin{center}
\includegraphics[width=\textwidth, trim= 0cm 0cm 0cm 0cm,clip]{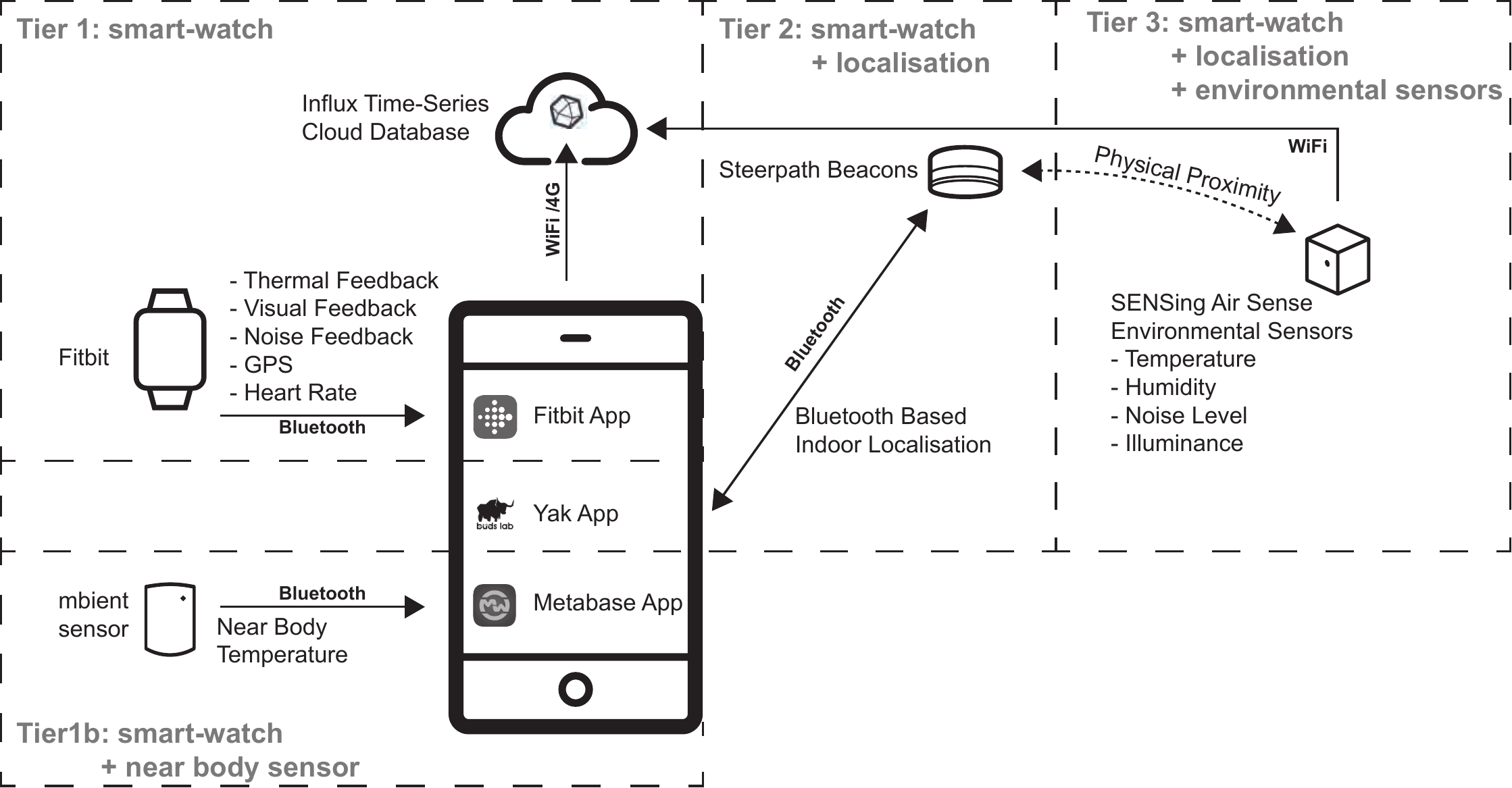}
\caption{Overview of the experimental deployment in the SDE buildings in four distinct tiers: 
Tier 1 is the base methodology which is production-ready for real-building deployment. It requires a smart watch with the cozie clock-face installed.
Tier 1b, is an extension to the base methodology by adding a temperature sensor to the watch. 
Tier 2 includes building-wide indoor localisation. In this experiment, Steerpath Bluetooth beacons were used, which communicate with the occupant's smartphone to determine the occupant's location.
Tier 3 merges the localised feedback points with environmental sensors in the same comfort zone as the occupant.
}
\label{fig:method}
\end{center}
\end{figure*}

The innovations outlined in this work as compared to the previously mentioned studies are:
\begin{itemize}
\item The hardware and software deployment methodology have a focus on practicality in scalable, field-based implementations. Experimental participants were only asked to wear a single smartwatch device and answer survey questions that utilise a relatively small amount of time. The focus was on testing a configuration that was easily applied in a real-world context. The modelling methodology was designed to capture as much signal as possible in the field setting without the ability to control and verify sensor proximity and accuracy consistently.
\item A series of pre-processing steps were developed to convert intensive longitudinal data into model input features that characterise the \emph{tendency of groups of people to have similar comfort preferences}, sometimes independent of the objective environmental factors such as temperature. A simple example of this concept is the commonly discussed, yet often anecdotal, person who seems always to need more cooling, even when the temperature is already low relative to the comfort zone. In this study, clustering was used to group people into \emph{comfort preference types} as an input feature to a preference prediction model.
\item This paper introduces and tests a simple form of a \emph{cold start} variant to the preference models that could be used to predict an occupant's preference with limited or no data about their preference history in a particular space or according to particular objective measurements such as temperature, humidity, or other factors. This model enables the deployment of the \emph{cozie} data collection methodology by a set of participants in a building and then the creation of prediction models that could accommodate future occupants regardless of whether they have worn a smartwatch in those spaces.
\item The process seeks to show that comfort-based preference prediction can be accurate even in the absence of environmental sensors if enough intensive longitudinal data has been collected from enough occupants. The context of this experiment was in relatively uncontrolled, field-based settings as opposed to laboratory conditions. 
\end{itemize}

\section{Methodology}
\label{ch:method}

To collect intensive longitudinal data in a field setting, the \emph{cozie} platform was built on the Fitbit smart watch\footnote{\url{https://www.fitbit.com/}} and various time-series database technologies. In this section, the details of this technology stack are explained in the context of a deployment on 30 test participants in buildings at the National University of Singapore (NUS) in the School of Design and Environment (SDE). The definition of an \emph{occupant} in this study was a test participant who wore the smartwatch, and a \emph{manager} as the person who coordinated the study. Thirty participants were recruited via an online form, were compared to the inclusion criteria for the study, and were on-boarded according to an approved ethics review application. Priority was given to participants who work full time in the SDE-related buildings on campus, and they were selected to maintain an even gender distribution. 

The technology used in the deployment of this study can be sub-sectioned into individual tiers as described in Figure \ref{fig:method}, with each level requiring additional resources to implement. For the experiment in SDE4, all tiers were incorporated.

\subsection{Tier 1: Smartwatch for micro-EMA}

Tier 1 is the core methodology presented in this paper, which uses the \emph{cozie} clock-face, as shown in Figure \ref{fig:cozie}. The occupants were asked to wear a Fitbit Versa smartwatch during daytime hours while on the NUS campus at the very least but were also welcome to wear the device for the entire duration of the study. Participants were asked to leave momentary assessment feedback on their comfort preferences at different points throughout the day on the watch face of the Fitbit device. Each time they responded to the survey, they were asked about their thermal, visual and aural preference using the options found in Figure \ref{fig:cozie}. Comfort preference was chosen as the feedback most applicable to the methodology due to a three-point scale that is most appropriate for frequent watch-based surveys. Preference surveys also provide more meaningful information by indicating how the occupant would want the environment to change as opposed to \emph{satisfaction} or \emph{sensation} survey types that only capture how the occupant feels. The participants were asked to answer the questions when they moved from one environment to another, which amounted to approximately 5-15 assessments per day. The smartwatch also prompted the occupants with a small vibration that requested feedback from them at different timed points in the day. This prompt only occurred during daytime hours when the subject was active. The momentary assessment took less than 15 seconds to complete. Throughout the experiment, the cumulative amount of time spent answering the momentary assessment was approximately 20-40 minutes.

Detailed documentation for using \emph{cozie}, along with the source code to an open-source repository, can be found on a GitHub repository\footnote{\url{https://github.com/buds-lab/cozie}}. The platform also can collect sensation, satisfaction and objective feedback such as clothing and activity levels. These features were added after the experiments outlined in this paper and were not used in this study.



\subsection{Tier 2: Indoor localisation}
Tier 1 is likely sufficient for experiments conducted in small office spaces. If only a few different zones exist, then an occupant's location could be quickly determined through a supplementary question in the question flow of the survey. However, in a large building, such as the SDE4 building where the outlined experiment was conducted, a more sophisticated indoor localisation system was required. The SDE4 building has six different floors, a gross floor area of around 8,500 square meters, and a large variety of different indoor environments. To determine an occupant's location in a building, 100 Bluetooth beacons and the Steerpath\footnote{\url{https://steerpath.com/}} platform were installed throughout the building. These beacons communicated with a custom-built smartphone application, called the \emph{Yak App} \citep{Abdelrahman2019}, to determine their location with a one-meter precision. The location data was then used to geo-fence the occupant within various zones of the building and was merged with the subjective preference feedback data in the cloud.


\subsection{Tier 3: Preference data convergence with environmental sensors}
Tier 3 included the deployment of 45 indoor and outdoor environmental quality (IEQ) sensors in the experimental context. This data collection tier was used to compare the results of the subjective feedback, with existing environmental models. The IEQ sensors were WiFi-connected and were deployed by the company SenSING\footnote{\url{https://sensing.online/}} as part of an installation of sensors campus-wide. These sensor kits measured temperature, humidity, noise level, and illuminance. At least one sensor device was installed in each zone of the building, and the data was pulled from an API and merged with the subjective preference data in the cloud. 

\subsection{Tier 1b: Strap-mounted sensor kit}

Tier 3b included a temperature sensor was used from mbient labs\footnote{\url{https://mbientlab.com/}}, which was attached to the watch through a custom three dimensional (3D) printed case. The design file for this case can be found online\footnote{\url{https://myhub.autodesk360.com/ue29ab3ac/g/shares/SH919a0QTf3c32634dcfe0a71457c4729699}}. The mbient device logged data locally, which was transferred to the cloud database at the end of the experiment.

\subsection{Occupant and room preference clustering}
This analysis included the hypothesis that the feedback of one of the occupants in such groups could be used to characterise the preferences of all group members for a particular space or set of conditions. In this step, the preference history of occupants was used to do a simple clustering-based segmentation step to group occupants according to their raw feedback preference tendencies. For example, occupants who more frequently indicated \emph{prefer cooler} as compared to a \emph{no change} would be grouped. This strategy was a simplified version of this type of clustering as it neglects other context-based variables (environmental and physiological measurements). This choice was made to keep the method feasible even in situations in which other measurements are not available. 

Given its widespread usage in related literature, occupant and room clustering is calculated using the k-means clustering algorithm with Euclidean distance, using the scikit-learn package\footnote{\url{https://scikit-learn.org/stable/}}. The features used for clustering were the ratio of votes of each feedback class value for each subject. For example, the ratio of \emph{prefer cooler} for a given participant, or room, would be calculated as follows: $\frac{\# \text{prefer cooler votes}}{\# \text{total votes}}$. This calculation is repeated for all types of feedback responses for thermal, light, and aural feedback. Then, the number of clusters was chosen to match the number of possible responses per type of feedback, this led to initially $k=9$, but given that there were no data points with \emph{prefer louder} responses, the clusters were merged into eight.


\subsection{Occupant comfort preference prediction}
The metric of comparison in this study was the prediction improvement of a machine learning model using added feature sets extracted from the intensive longitudinal preference data. This structure matches implementation-based environmental comfort studies outlined in the literature that showed the predictive improvement of additional data \citep{Liu2019-pi, Aryal2019-bx}. This approach can be compared to more controlled, lab-based methods that seek to isolate variables and individually test their influence. 

The prediction problem translates to predicting the right class value or, in this case, the preference feedback response, at the given feature values. A random forest classifier from the scikit learn package was chosen to handle this comfort prediction. Random forest classifiers have been proven to have the highest accuracy at predicting personal comfort in one previous study \citep{kim2018paradigm} and is one of the best performing of other recent studies \citep{Liu2019-pi, Aryal2019-bx, Luo2020}. The decision was made to focus on the implementation of a single model type that has been proven effective and is straightforward to use based on documentation and ease-of-tuning. With this in mind, we fixed the hyper-parameters for the random forest classifier to 1000 number of trees, Gini criterion for node splitting, and two minimum samples per split.

Additionally, the prediction problem was divided into an individual and a grouped prediction task. The former refers to a model developed specifically for a given occupant, using only parts of its data to train a model and test it on its remaining data. On the other hand, the latter approach consists of combining all occupants' training data subsets with training a model and testing it on all the occupants' remaining data.

The data of each occupant was split into a 60:40 train test set based on time. That is, the first 60\% of votes from each occupant was used in their training set, and the remaining 40\% was used for testing. The sets were split by time to prevent the scenario of future data being used to predict the past. For the grouped model, all the occupants' training sets (60\% of each occupant's data) was used as one training set, and the remaining 40\% of each occupant's data was combined with being used as one test set.

A primary component of the method was to test the ability for various feature sets to influence the prediction power of the random forest model. The method used six combinations of these feature sets to test the influence each has in the predictive capability of the overall model. The following is an overview of these feature categories developed for testing:

\begin{itemize}
\item \textbf{Time} was created through feature engineering the time stamp of when an occupant gave feedback. This feature was a cyclical representation of the hour of the day and day of the week. This simple feature type detects if certain cyclical habits or components have a role in preference prediction and was included in all scenarios.
\item \textbf{Environmental Sensors} were features extracted from measurement data from lighting (lux level), noise (dB level), temperature (deg. Celsius), and relative humidity (RH\%) measurement. These variables were collected from the IEQ sensors that were closest spatially and temporally to an occupant when they gave feedback.  
\item \textbf{Near Body Temperature} was a feature created from the temperature sensor mounted on the smartwatch strap that had temporal proximity to the time-stamp of when the occupant gave feedback.
\item \textbf{Heart Rate} was collected from the Fitbit smartwatch device as an instantaneous value collected when the occupant gave feedback.
\item \textbf{Room} was a feature that was encoded to a numerical preference type based on the history of feedback in the room in which the survey was taken. This feature was designed to increase the prediction accuracy by complimenting data from rooms of similar comfort profiles. For example, if an occupant only works from their office, the model will still be able to accurately predict how that occupant may feel in other rooms that have a similar comfort profile to their office. 
\item \textbf{Preference History} features are similar to the Room features. These features use the ratio of responses of each type (thermal, visual, and aural) that were calculated for each user.  This ratio was only calculated for the responses of \emph{prefer cooler}, \emph{prefer warmer}, \emph{prefer dimmer}, \emph{prefer brighter}, \emph{prefer quieter}, and \emph{prefer louder}. E.g., the ratio of response of \emph{prefer cooler} responses of a given occupant is calculated the following way: $\frac{\# \text{prefer cooler votes}}{\# \text{total votes}}$.
\end{itemize}



Model classification results were calculated using the F1-micro scores (as shown in Equation \ref{eq:f1}) which were equivalent to accuracy in the a multi-class classification problem by calculating \texttt{precision} and \texttt{recall} averaged across all classes, i.e., subjective thermal comfort response value. As the objective was to provide a comparison among different feature sets with a standard metric, F1-micro was chosen due to its usage for benchmarking different aspects of the modelling pipeline in thermal comfort datasets \citep{Luo2020}. 


\begin{equation}
F1 = 2 \cdot \frac{precision \cdot recall}{precision + recall}
\label{eq:f1}    
\end{equation}

\begin{figure}[h!]
\begin{center}
\includegraphics[width=0.45\textwidth, trim= 0cm 0cm 0cm 0cm,clip]{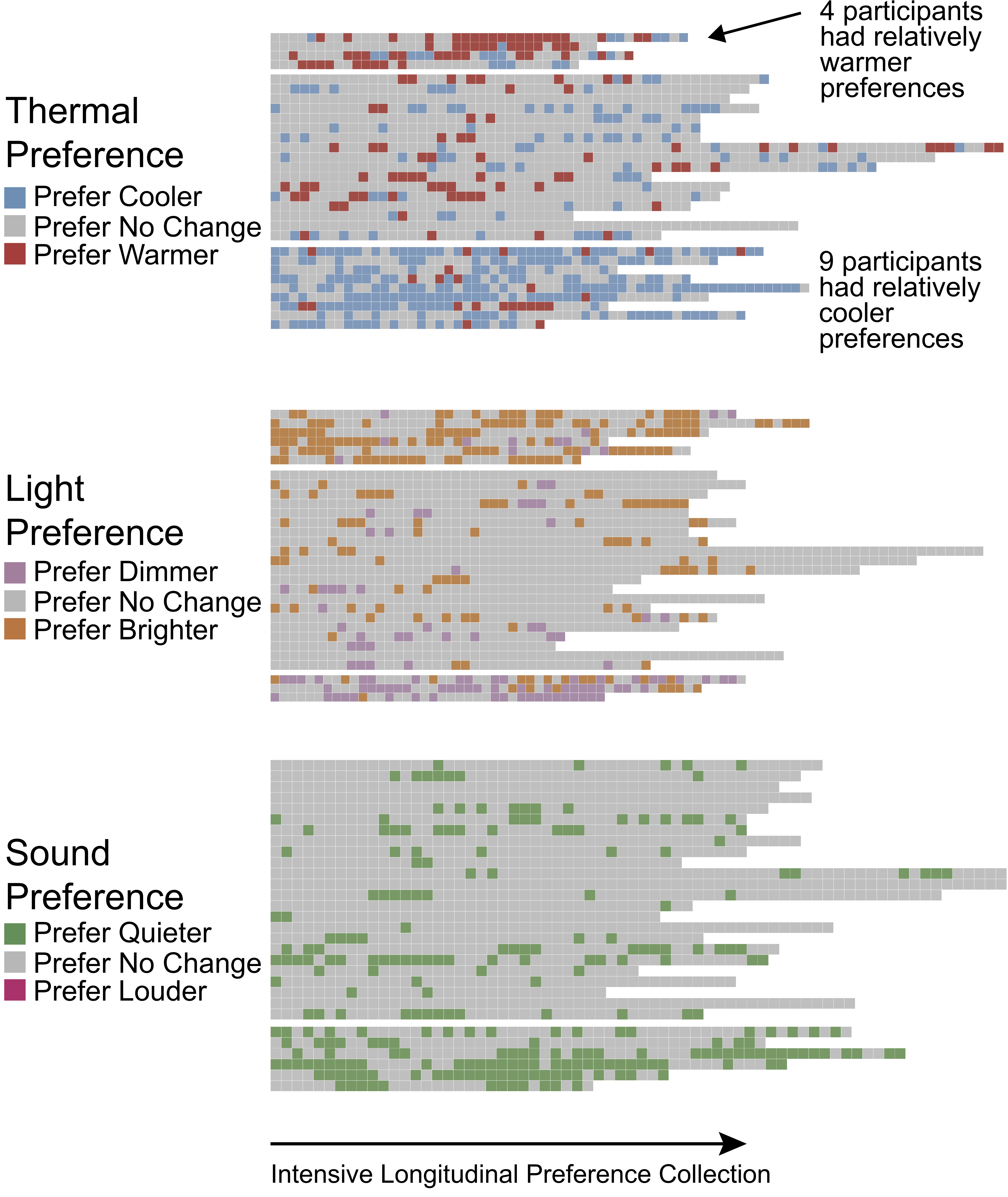}
\caption{Overview of the intensive longitudinal data collected from the occupants according to the three categories: \emph{Prefer warmer} (red) and \emph{Prefer cooler} (blue) for thermal, \emph{Prefer Brighter} (orange) and \emph{Prefer Dimmer} (purple) for lighting, and \emph{Prefer Quieter} (green) for acoustics. Each row is an occupant and each box in that row shows that occupant's feedback answers collected sequentially. The visualisation is diagrammatic in that vertical alignment of the boxes between different occupants does not imply identical time stamps.}
\label{fig:longclustering}
\end{center}
\end{figure}

\begin{figure*}
\begin{center}
\includegraphics[width=\textwidth, trim= 0cm 0cm 0cm 0cm,clip]{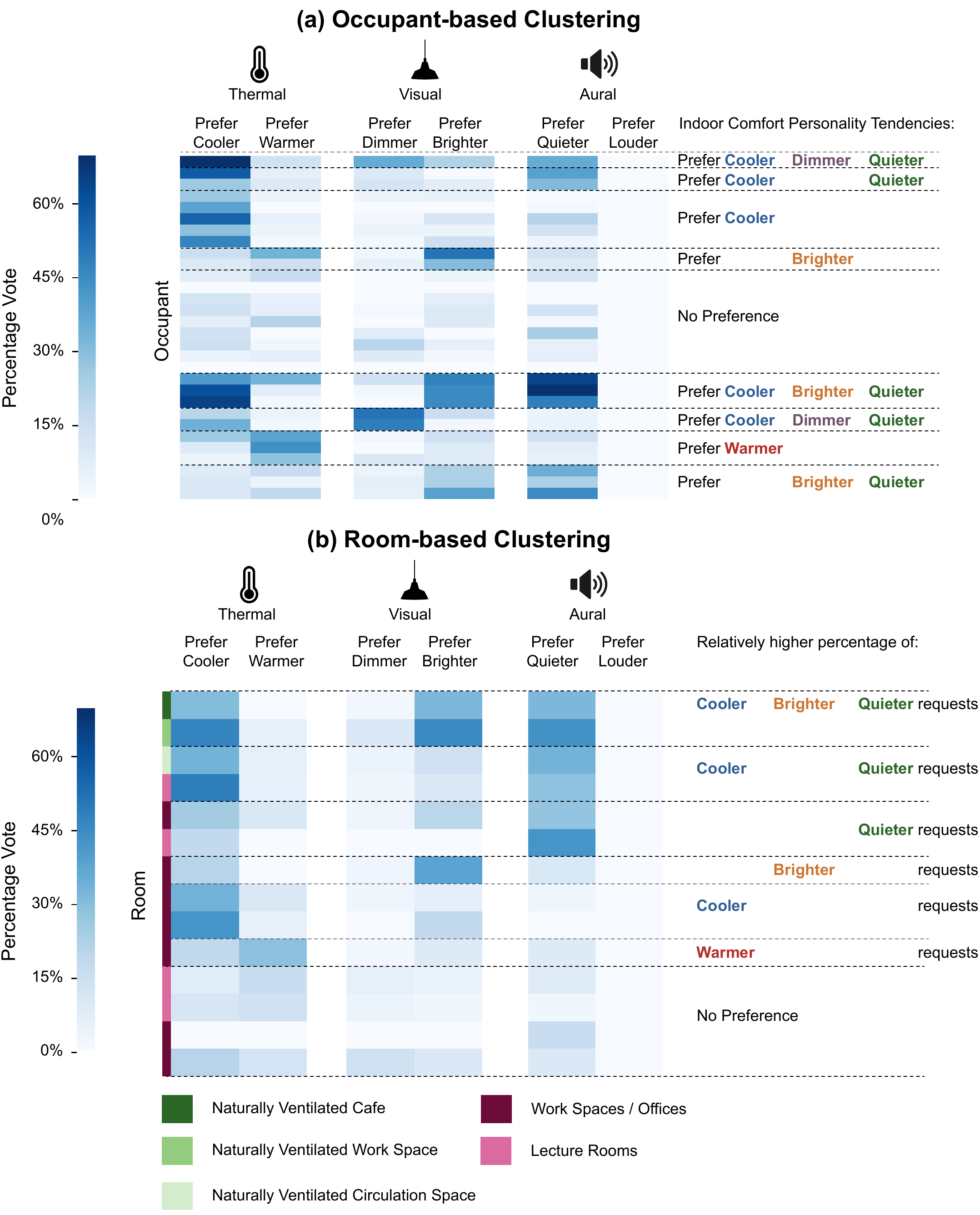}
\caption{K-means clustering of preference tendencies quantified by average number of votes by occupant (a), and by room (b) within the test building. Each row presents the percentage of votes that fell into a respective preference. Dark colours in cells indicate higher preference.}
\label{fig:clustermap}
\end{center}
\end{figure*}

\begin{figure*}
\begin{center}
\includegraphics[width=\textwidth, trim= 0cm 0cm 0cm 0cm,clip]{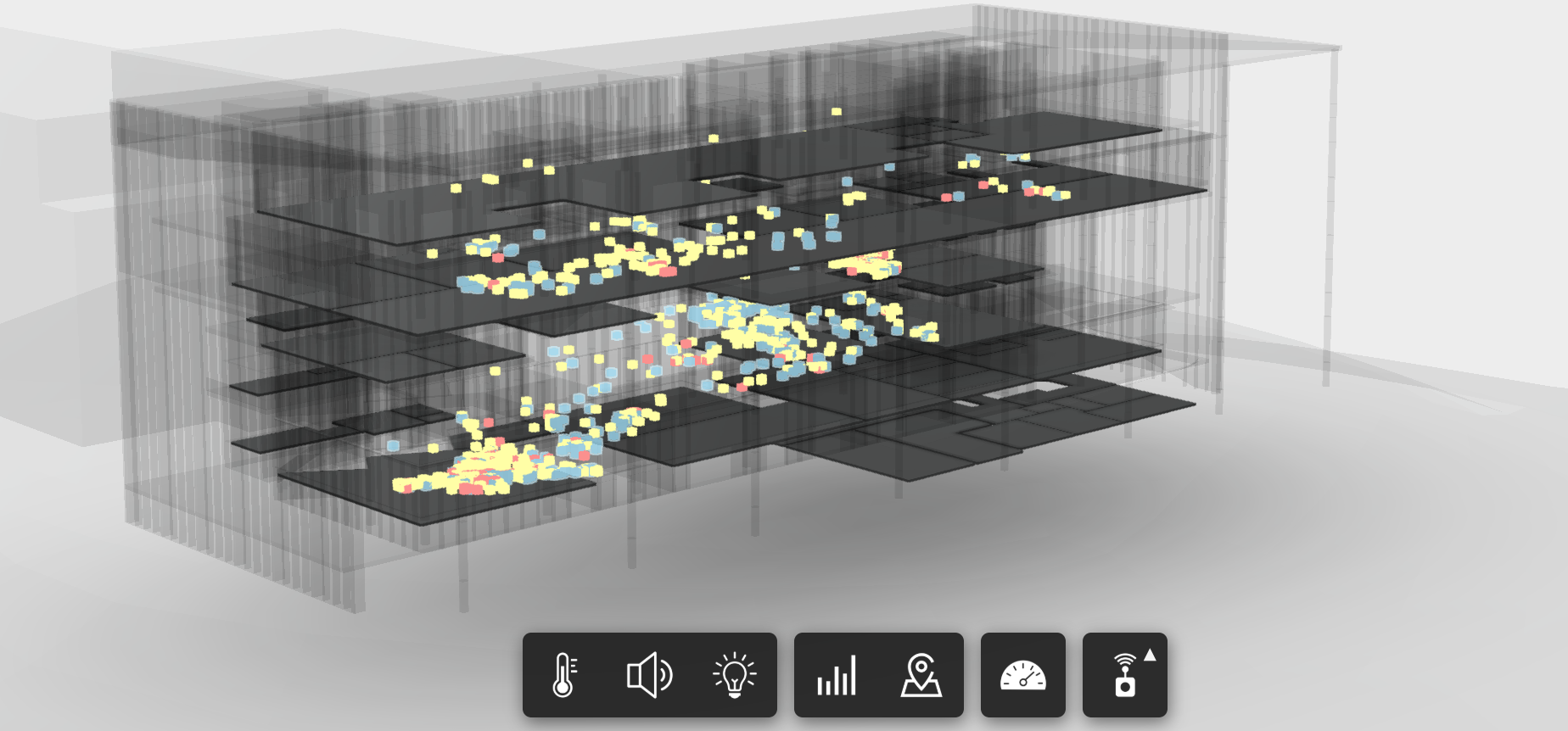}
\caption{Interactive visualisation of the data collection in the SDE4 building that highlights the spatial distribution of subjective preference data for thermal comfort in three dimensions. In terms of preference feedback, the blue dots indicate \emph{prefer cooler} responses, the yellow dots are \emph{no change}, and the red dots are \emph{prefer warmer}.}
\label{fig:sde4demo}
\end{center}
\end{figure*}

\begin{figure*}[h!]
\begin{center}
\includegraphics[width=\textwidth, trim= 0cm 0cm 0cm 0cm,clip]{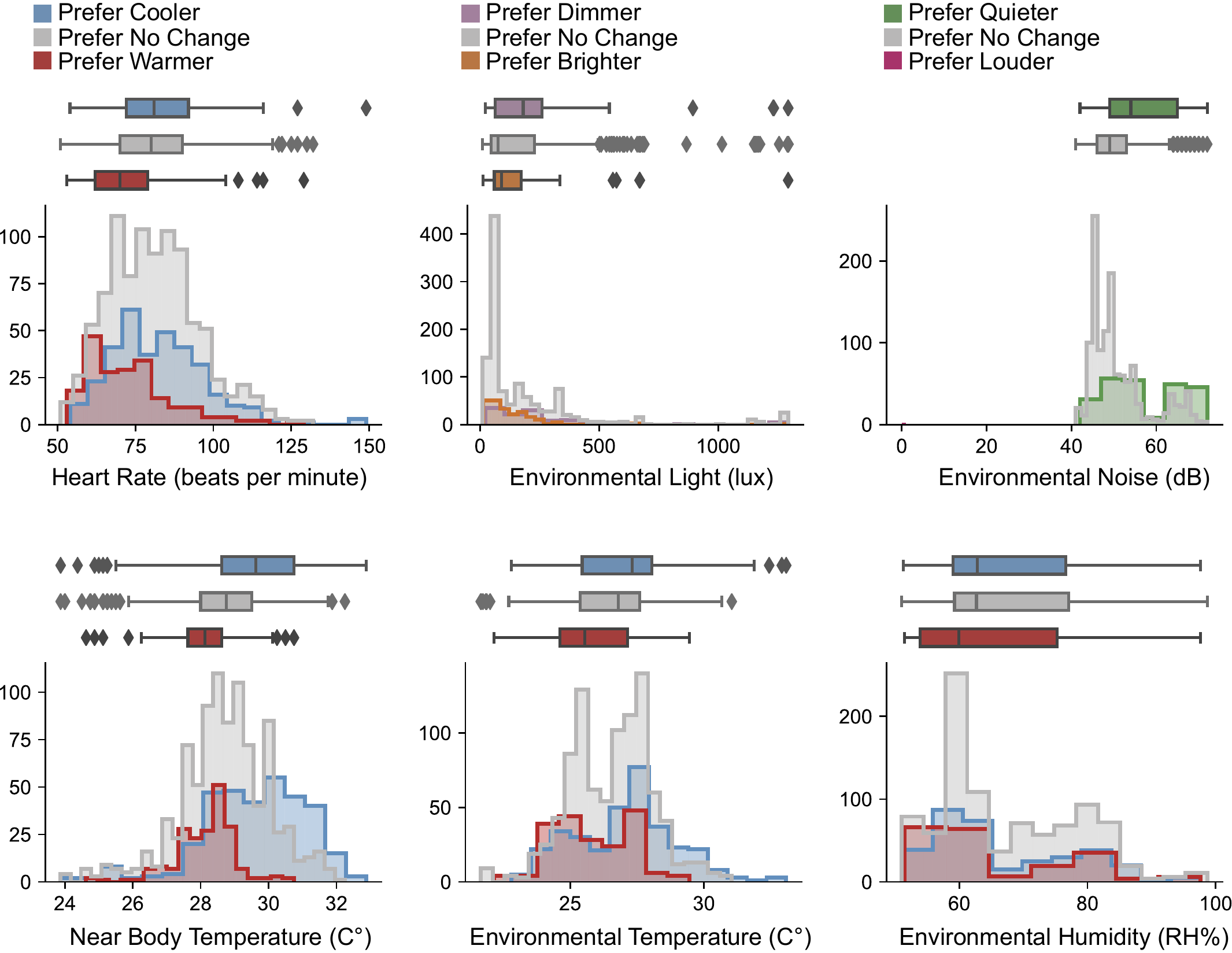}
\caption{Distribution of sensor data by preference vote. While trends can be observed many feedback votes overlap for the same environmental or physiological measurement. This was possibly due to the different comfort tendencies as shown in Figure \ref{fig:clustermap} or numerous other variables described in Figure \ref{fig:variables} that are not accounted for. Near body temperature and noise appear to have the most distinct differentiation.}
\label{fig:sensor_data}
\end{center}
\end{figure*}

\section{Results}

\label{ch:results}
The results presented in this section are complemented with an interactive web application\footnote{\url{https://sde4demo.herokuapp.com/}} and interactive code\footnote{\url{https://github.com/buds-lab/humans-as-a-sensor-for-buildings}} which enables the reader to regenerate all the plots. During a two week collection time of 30 participants, 4,378 comfort preference votes were collected, which is 146 data feedback points per person on average. From this set, 1,474 data points were successfully localised to building environmental sensors. To allow for comparison with those data, this subset was used for analysis and machine learning in the following sections.

\subsection{Grouping comfort preference tendencies}
\label{ch:tier1result}

Figure \ref{fig:longclustering} illustrates an overview of the intensive longitudinal preference history data for each person according to the three preference categories. These feedback responses were only those collected in the SDE4 building, and a maximum number of 75 votes is shown. A simple clustering step was applied in this figure to represent the segmentation according to each preference category on its own. This visualisation shows how this simplified clustering step captured the \emph{tendency} of an occupant to lean more towards one feedback response over the others. This segmentation was independent of the environmental parameters of the spaces to maintain the simplicity of the approach. The subsequent modelling steps were designed to test the effectiveness of doing this type of simplified segmentation.

Figure \ref{fig:clustermap}a is an aggregated representation of the segmentation process for each occupant, this time with all three preference categories being used in the clustering process. This figure summarises each occupant as a row of data, and the colour of the box represents the percentage of votes given to a particular preference category, where dark colours indicate higher preference. These clusters provided segmentation of the users according to their preference tendency types that were used in the preference models. Even in a sample size of 30 occupants, there were varying comfort tendencies present, which complemented the concept of a personal comfort model tested by Kim et al. \citep{kim2018personal}. This clustering step provided the foundation for the creation of the \emph{individual versus grouped} models used in the prediction step.

\subsection{Tagging the spatial context with preference feedback}
While the subjective feedback highlighted varying comfort tendencies within a building, localisation also enabled the characterisation of preference tendencies in certain zones. Figure \ref{fig:clustermap}b presents each room as a row, where the colour of each cell represents the percentage of a preference vote given for a particular room. The utilisation of k-means clustering once again enabled the splitting and labelling of these zones, this time by the tendency for different comfort preferences to be left by occupants in those spaces. This result firstly served as an overview for facility managers to understand the office spaces they manage, and take action to improve upon the comfort. A visualisation of the subjective thermal preference data can be found in Figure \ref{fig:sde4demo} and online\footnote{\url{https://sde4demo.herokuapp.com/}}.

\subsection{Correlation with indoor environmental quality variables}
One standard aspect of environmental comfort studies is the comparison of feedback to objective environmental measurements. For the data collected in this study, standard distribution plots of the environmental sensor data are summarised in Figure \ref{fig:sensor_data}. Intuitive insight in the data can be observed, such as the absence of \emph{prefer brighter} votes after an illuminance threshold of 250 lux. Nevertheless, there was a significant overlap between classes for each of the environmental parameters, which were likely attributed to the numerous unmeasured variables described by Figure \ref{fig:variables}, and the varying comfort tendencies shown in Figure \ref{fig:clustermap}. This result reinforces the evidence that environmental measurements are not descriptive enough to characterise a person's preferences, which results in poor prediction as found in previous studies \citep{cheung2019analysis}.

\subsection{Predicting field-based indoor preference using intensive longitudinal data}

In this section, the time-series feedback was used to predict comfort satisfaction. Figure \ref{fig:feature_compare} shows a comparison of the various models built with the feature sets and process outlined in Section \ref{ch:method}. The individual comfort model uses the occupant's training data for prediction, while the grouped comfort model uses the input data for the groupings outlined in Figure \ref{fig:clustermap}. The top of Figure \ref{fig:feature_compare} shows a table in which each row represents the feature set that was used to train the model in that column. 

Several insights were evident from this modelling analysis. Firstly, there were only small differences in the F1 scores between the different feature sets for the visual and aural preference models. These models, in general, had higher F1 scores than thermal preference prediction. Aural preference prediction had the highest F1 score, which was intuitive since it became a binary classification challenge due to the lack of \emph{prefer louder} feedback responses. 

Thermal preference prediction had more diversity across the feature sets tested as compared to the other preference categories. Merely using the conventional time-series and environmental sensor features had the lowest F1 score. Adding the physiological attributes of heart rate and near body temperature provided marginal improvements. The best thermal preference model used the physiological, room, and preference history features while excluding the environmental sensor data. 

For all three preference categories, the grouped comfort model performed better than the individual version. Participants with similar comfort preferences became clustered together, thus increasing the training dataset for that particular occupant type. This result showed the impact that assigning a variety of \emph{peer groups} can have on preference prediction.

\subsection{Cold-start comfort preference prediction}

\begin{figure*}
\begin{center}
\includegraphics[width=\textwidth, trim= 0cm 0cm 0cm 0cm,clip]{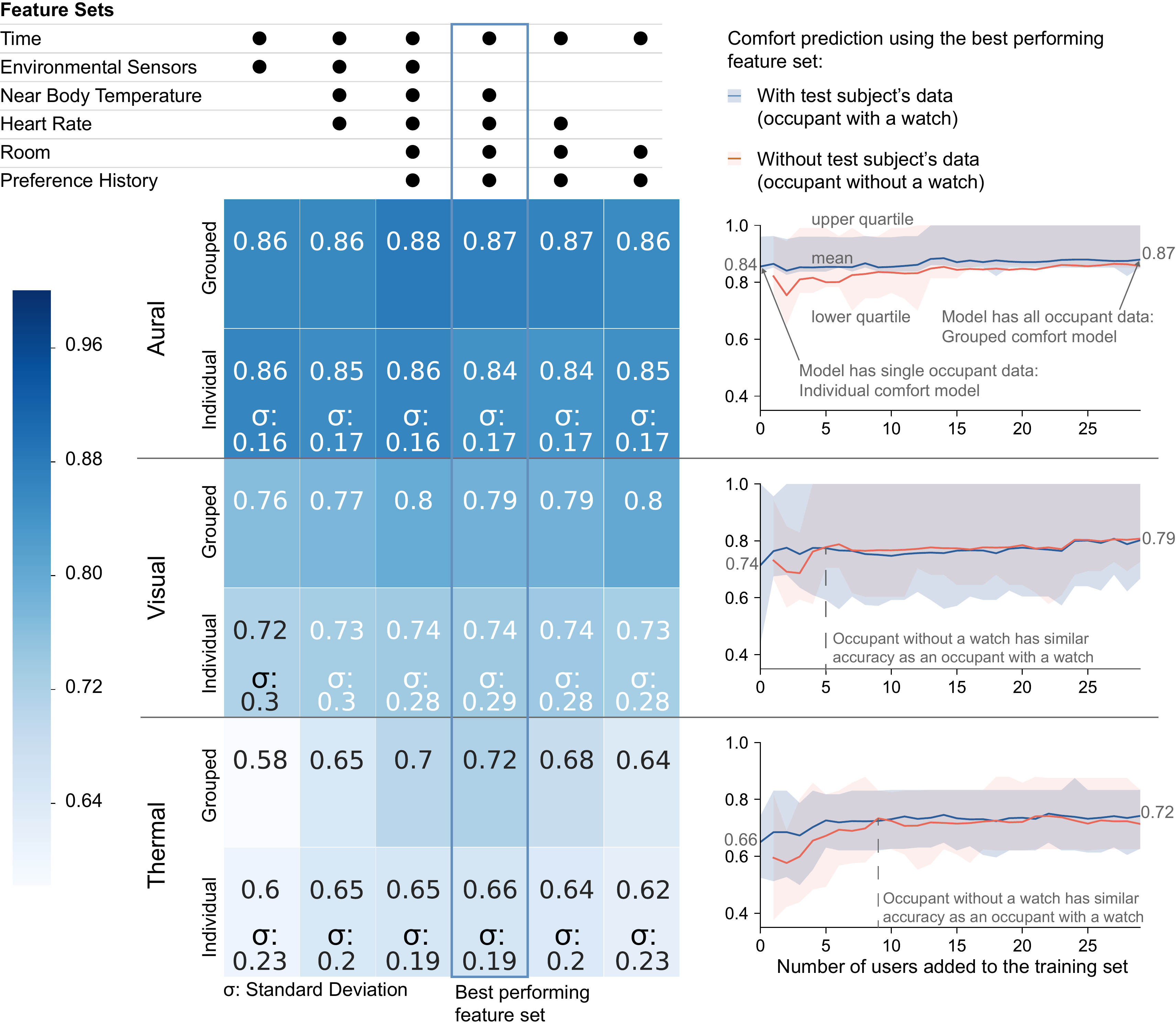}
\caption{Left: Comparison of prediction F1-micro-score between grouped and individual comfort models using data from different feature sets. The feature set that excluded environmental sensor data for the thermal model had the highest F1-score, while minimal differences in F1-score were noted between feature sets of the visual and aural models. Right: The accuracy in predicting the comfort of an individual as further participants are added to the training set. The blue line includes the test participants training set in the training data, and the orange line excludes the test the participants training data meaning that it depends on crowdsourced feedback from other occupants.}
\label{fig:feature_compare}
\end{center}
\end{figure*}

The success of the preference models using grouping allowed for models that can predict an occupant's preferences without their own personal data present. This scenario was labelled as a \emph{cold-start} situation as it emulates when an occupant doesn't wear a watch to collect data in a particular building, but relies on crowdsourced data from peers that have a similar comfort preference. The line graphs to the right of Figure \ref{fig:feature_compare} show the results of this type of analysis. They illustrate the number of occupants required to sufficiently crowdsource the data for an average occupant for each of the preference categories. The orange line represents an ordinary person who doesn't wear a smartwatch, whereas the blue line is a smartwatch owner who is regularly giving feedback. In this study, nine and five users were sufficient on average to crowdsource the thermal and visual comfort prediction respectively to the same accuracy as a user wearing a watch.

\begin{figure*}
\begin{center}
\includegraphics[width=\textwidth, trim= 0cm 0cm 0cm 0cm,clip]{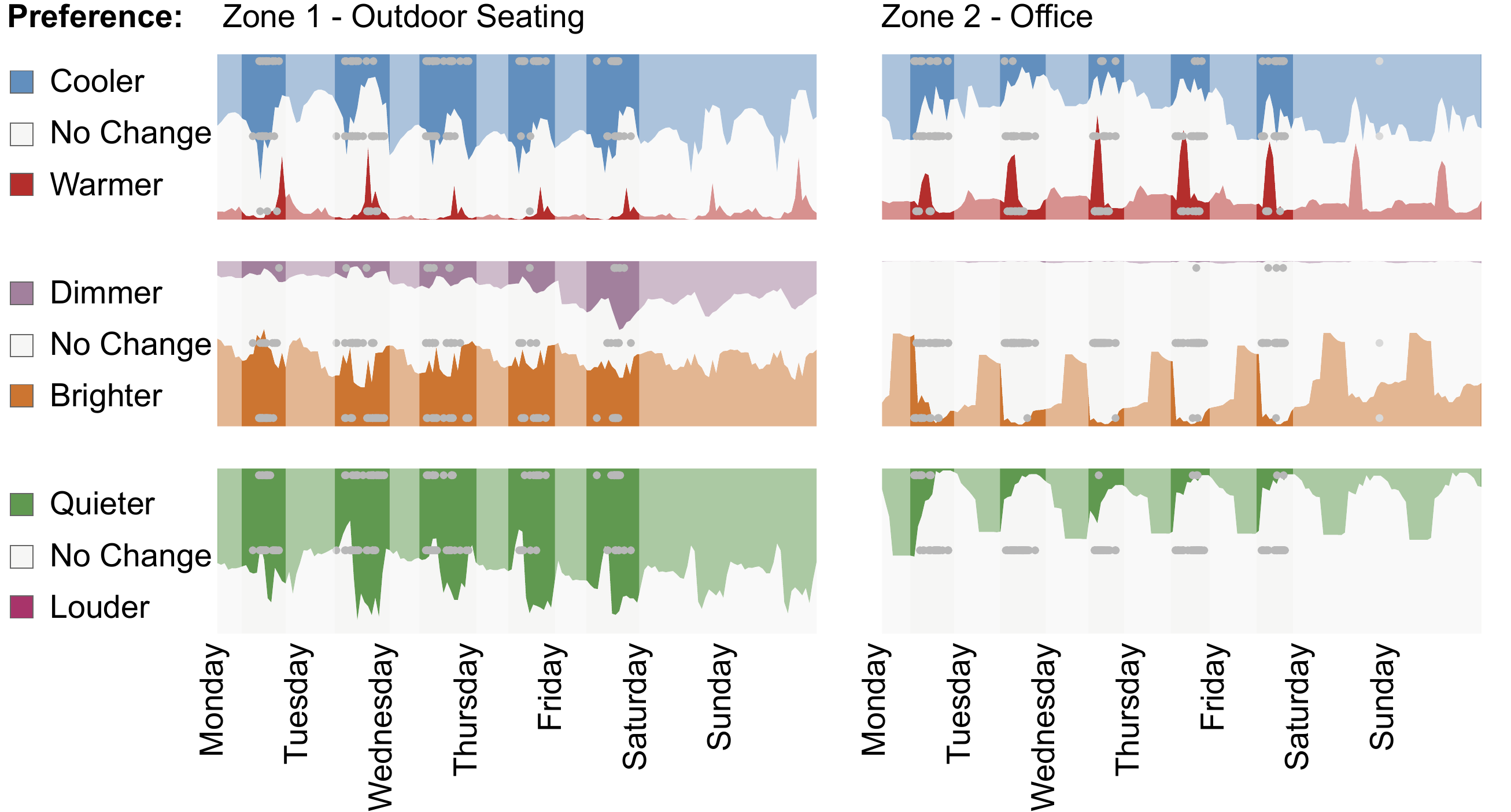}
\caption{Comfort prediction for two zones for an average occupant over a week. The grey circles indicate votes that were given for each category, and the shaded-out sections indicate times where no data were present. These time-series predictions can be used to detect anomalies, such as the mid-day peak for a warmer preference for the \emph{office}, or the general discomfort in the \emph{outdoor seating} area. Note that there was an absence of data in these zones between the hours of 22:00 - 7:00, and on the weekend. This lack of data caused inaccurate predictions as seen in the square-shaped peaks in the \emph{office}.}
\label{fig:comfort_forcast}
\end{center}
\end{figure*}

\subsection{Predicting continuous comfort preference without sensors}

Since the preference feedback in this methodology was at a much higher high-frequency than a typical survey or occupants acting on the thermostat, this study had preference data with relatively high temporal and spatial diversity. The random forest classifier was used to predict comfort preference based on a time-stamp input for each zone to create a continuous prediction over time. This approach emulates the concept of using human feedback as a type of sensor. Figure \ref{fig:comfort_forcast} illustrates the prediction of two different zones, an office and an outdoor space, for a typical week using this model output. First, one can see that the \emph{office} was generally a comfortable space, while the \emph{outdoor seating} had an overall higher preference for cooling. Time-dependent fluctuations show how the model was able to predict comfort preference for different parts of the day or days of the week. The \emph{office} had a peak of warmer preference around mid-day.  Finally, it was observed how the model, often inaccurately, tried to predict comfort at times where no data is present. The square peaks in the \emph{office} for aural and visual prediction between the hours of 22:00 and 7:00 were due to an absence of data to accurately predict during these times.

\section{Discussion}

\label{ch:discussion}

The results of this implementation showed the potential of collecting intensive longitudinal feedback from occupants in the built environment. This approach revealed that the deployment and implementation of such a methodology were effective, and comfort models for visual, aural, and thermal comfort can have similar performance to sensor measurements. The key focus in this section is to discuss the practical uses and limitations of the proposed methodology.

\subsection{Practical Application of Intensive Longitudinal Data in Industry}
At the foundation of the method, the creation of more significant amounts of occupant feedback information in the form of preferences was successful. The utilisation of these type of high-frequency subjective feedback data has potential for building evaluation and occupant comfort optimisation. It changes the paradigm in which facility managers operate a building. For example, instead of saying \emph{light levels are below the comfort threshold in Office-1}, the new conclusions could state that \emph{a higher frequency of prefer brighter votes are recorded in Office-1}. Furthermore, due to the high-frequency sampling rate provided by the micro ecological momentary assessment methodology, these periods of discomfort can also be mapped to particular times, and certain groups of people. The time series comfort profiles could also serve as input data for occupant-centric-control efforts of building systems which can then optimise for human comfort and energy optimisation.

\subsubsection{Post-occupancy evaluation, commissioning, and sensor calibration}
A focus on post-occupancy evaluation could be a key target for this type of data collection. In this scenario, a particular sample of non-transient occupants of a recently constructed building would be given smartwatches and asked to wear them for 2-4 weeks. These data could then be used to supplement the systems installed to characterise whether there are \emph{blind spots} in terms of sensors not picking up comfort-influencing phenomenon that is not being measured. For example, it's rare to measure mean radiant temperature in most buildings; therefore \emph{hot spots} might exist that are undetectable by thermostats and might be a result of inadequate shading or control of shading systems. To adapt the presented methodology to this context is straightforward as the two-week time frame of the experiment is similar. The current method involved asking each participant to wear the smartwatch until 100 data points were recorded. In a real-world setting, a similar approach could be deployed. At that point, perhaps the occupant could choose to return the device and rely on the data of co-workers for comfort prediction or continue to wear it and help crowdsource the data for others. It is strongly recommended that these deployments use automated indoor localisation to put the feedback in the spatial context without user intervention.


\subsubsection{Potential for spatial recommendation systems and impact on activity-based workspace design}
A less typical application for intensive longitudinal data might be the development of spatial recommendation engines for occupants in activity-based workspaces. In these spaces, an occupant doesn't have a constant workspace but instead finds a space that matches their immediate needs. This paradigm could prove to be an integral part of future working style, especially in light of social distancing due to global pandemics such as COVID-19 that forces a less conventional spatial working arrangement. This recommendation engine might work in a way that an occupant's comfort tendencies could be matched with the comfort zone of the building. For example, those that prefer warmer spaces can be recommended to work in areas that have a higher percentage of \emph{prefer cooler} votes. Previous work in this direction showed progress using a platform known as Spacematch \citep{Sood2020-vz}. Smart watch-based longitudinal feedback could enhance the model development process for this type of platform.

The aspect of testing group-based models in this study is essential for this context as building owners can't expect all occupants to be willing to wear or use devices. And those that do agree will likely have a limited amount of patience for giving feedback over long periods. This paper tested the ability to cluster occupants such that it was not necessary for everyone in an office to wear a smartwatch. The only requirement for this type of system to work is that each new employee would wear a smartwatch during a two-week data collection phase, which is sufficient to build their comfort preference tendency history as described in Section \ref{ch:tier1result}. The experiment also showed that not everyone in an office space needed to be using the smartwatch application all the time. In this particular experiment, six occupants were sufficient, on average, to crowdsource the prediction for the remaining 24. This value is not generalisable amongst all buildings and would change depending on the different comfort tendencies the building occupants might present. The higher the variation in comfort preferences, the greater the number of the occupants needed to crowdsource the data.


In terms of office space design, the collection of intensive longitudinal preference data could facilitate floor plan design decisions. Understanding the breakdown of comfort needs according to the tendencies of the occupants would enable architects to design or retrofit buildings with different comfort zones to match the different types of people. For example, if the zones with more cooling were popular and being used to their capacity, then the floor plans or systems control could respond to by creating more spaces of that type to increase the probability that a person feels comfortable.

\subsubsection{Integration into building control systems}
Intensive longitudinal data has the opportunity to influence the control and automation systems of buildings through the use of preference feedback data in the control logic. Most building control systems rely on optimising a set-point temperature that is considered comfortable for the average occupant or comfort standard \citep{Enescu2017}. In that scenario, discomfort instead of comfort is evaluated as the current difference of the environment thermostat and the HVAC system set-temperature \citep{Barrios2017}, occupancy density estimation, or via more traditional ways such as PMV \citep{Park2018}. While some of these approaches have dealt with single-occupant offices or Personal Comfort Systems (PCS), there is a distinction between controlling the actual HVAC system and allowing the occupant to control their immediate space. PCS systems are those that locally condition the occupant independent of the centralised HVAC system \citep{Zhang2015}. The intensive longitudinal data and the models developed in this study could help the controls field take the next step forward in occupant-centred building controls through the use of reinforcement learning \citep{Park2019-nm}. The feedback mechanism in reinforcement control is generally the standard occupant-building interface such as switch or thermostat \citep{Park2019-qm}. Intensive longitudinal data could be used to enhance that interaction by focusing on finding the motivations of those control actions. This work is a strong focus of the occupant-centric building operations in the IEA Annex 79 project \citep{OBrien2020-ns}.

\subsection{Prediction models are only as good as the training data}
The primary limitation of the presented approach was that it would only work where data were present. As seen in Figure \ref{fig:comfort_forcast}, there were errors in the prediction when data were absent, i.e., no historical data collected at similar time windows such as the middle of the night. Furthermore, since there was a reliance on other subjects' historical preferences, i.e., crowdsourcing preferences, to evaluate environments, an office space that was rarely used would have a poor prediction of occupant comfort. Classical comfort models based on sensor data do not have this issue as spaces that are not used can still be characterised by the measured data. Furthermore, this particular study was conducted in Singapore, which doesn't have seasons and has minimal variability in temperature. For seasonal countries, the day of the year would be an added feature that may take up to a year worth of data to train. Further work could investigate the opportunity of using sensor data to characterise a space and then continuously refine the comfort prediction by crowdsourcing the occupants' preference on said space.

\section{Conclusions}

\label{ch:conclusion}
This paper presents how micro ecological momentary assessments of subjective comfort can generate sufficiently large intensive longitudinal data for occupant comfort prediction and enhancement that can supplement objective environmental sensor data, and empirical comfort models. Results of an implementation of the platform on 30 occupants showed the segmentation and variation of indoor occupant comfort tendencies and highlighted the shortcomings of one-size-fits-all comfort models that are commonly applied in real buildings. Furthermore, the use of a smartwatch enabled data collection at a sufficient frequency to build time-series models of indoor spaces. These models could be used to detect building anomalies, serve as building data for subjective driven building control, or be used to recommend spaces that best match the comfort preference tendencies of each occupant. The optimum technological set-up uses a smartwatch for subjective data collection, combined with a method for localising an occupant in the building. This localisation may be achieved by asking the occupant directly through the smartwatch, or through Bluetooth or WiFi signals.



\section*{Author contribution statement}
PJ: hardware, software, infrastructure development, experimental design, implementation and lead author of the paper; MQ: software, infrastructure development, data analysis, machine learning lead and author of the paper; MA: software, infrastructure development, and author of the paper; CM: funding, project leadership, experimental design, the corresponding author of the paper.

\section*{Funding}
The Singapore Ministry of Education (MOE) (R296000181133 and R296000214114) and the National University of Singapore (R296000158646) provided support for the development and implementation of this research.

\section*{Acknowledgements}
This research contributes to the body of work for the International Energy Agency (IEA) Energy in Building and Communities (EBC) Annex 79 - Occupant-Centric Building Design and Operation.

\section*{Data availability statement}
Segments of the raw data and analysis code used for this study are available in an open-access Github repository that includes further documentation\footnote{\url{https://github.com/buds-lab/humans-as-a-sensor-for-buildings}}. 

\bibliographystyle{model1-num-names}
\bibliography{references}

\begin{thebibliography}{90}
\expandafter\ifx\csname natexlab\endcsname\relax\def\natexlab#1{#1}\fi
\providecommand{\bibinfo}[2]{#2}
\ifx\xfnm\relax \def\xfnm[#1]{\unskip,\space#1}\fi
\bibitem[{Frontczak et~al.(2012)Frontczak, Schiavon, Goins, Arens, Zhang, and
  Wargocki}]{frontczak2012quantitative}
\bibinfo{author}{M.~Frontczak}, \bibinfo{author}{S.~Schiavon},
  \bibinfo{author}{J.~Goins}, \bibinfo{author}{E.~Arens},
  \bibinfo{author}{H.~Zhang}, \bibinfo{author}{P.~Wargocki},
\newblock \bibinfo{title}{Quantitative relationships between occupant
  satisfaction and satisfaction aspects of indoor environmental quality and
  building design},
\newblock \bibinfo{journal}{Indoor air} \bibinfo{volume}{22}
  (\bibinfo{year}{2012}) \bibinfo{pages}{119--131}.
\bibitem[{Sakellaris et~al.(2016)Sakellaris, Saraga, Mandin, Roda, Fossati,
  De~Kluizenaar, Carrer, Dimitroulopoulou, Mihucz, Szigeti
  et~al.}]{sakellaris2016perceived}
\bibinfo{author}{I.~Sakellaris}, \bibinfo{author}{D.~Saraga},
  \bibinfo{author}{C.~Mandin}, \bibinfo{author}{C.~Roda},
  \bibinfo{author}{S.~Fossati}, \bibinfo{author}{Y.~De~Kluizenaar},
  \bibinfo{author}{P.~Carrer}, \bibinfo{author}{S.~Dimitroulopoulou},
  \bibinfo{author}{V.~Mihucz}, \bibinfo{author}{T.~Szigeti}, et~al.,
\newblock \bibinfo{title}{Perceived indoor environment and occupants’ comfort
  in european “modern” office buildings: The officair study},
\newblock \bibinfo{journal}{International journal of environmental research and
  public health} \bibinfo{volume}{13} (\bibinfo{year}{2016})
  \bibinfo{pages}{444}.
\bibitem[{Fanger et~al.(1970)}]{fanger1970thermal}
\bibinfo{author}{P.~O. Fanger}, et~al.,
\newblock \bibinfo{title}{Thermal comfort. analysis and applications in
  environmental engineering.},
\newblock \bibinfo{journal}{Thermal comfort. Analysis and applications in
  environmental engineering.}  (\bibinfo{year}{1970}).
\bibitem[{De~Dear and Brager(1998)}]{de1998developing}
\bibinfo{author}{R.~De~Dear}, \bibinfo{author}{G.~S. Brager},
\newblock \bibinfo{title}{Developing an adaptive model of thermal comfort and
  preference},
\newblock \bibinfo{journal}{ASHRAE Transactions, Vol 104}
  (\bibinfo{year}{1998}).
\bibitem[{Asadi et~al.(2016)Asadi, Hussein, Palanisamy, Conference, Comfort,
  Cumberland, Use, Environmental, Engineering, Mesiano, Sulaiman, Kamaruzzaman,
  Ramos, Dedesko, Siegel, Gilbert, Stephens, Asadi, Mahyuddin, Shafigh, Vieira,
  da~Silva, and de~Souza}]{Asadi2016}
\bibinfo{author}{I.~Asadi}, \bibinfo{author}{I.~Hussein},
  \bibinfo{author}{K.~Palanisamy}, \bibinfo{author}{W.~Conference},
  \bibinfo{author}{M.~Comfort}, \bibinfo{author}{R.~Cumberland},
  \bibinfo{author}{E.~Use}, \bibinfo{author}{C.~Environmental},
  \bibinfo{author}{M.~Engineering}, \bibinfo{author}{V.~Mesiano},
  \bibinfo{author}{R.~Sulaiman}, \bibinfo{author}{S.~N. Kamaruzzaman},
  \bibinfo{author}{T.~Ramos}, \bibinfo{author}{S.~Dedesko},
  \bibinfo{author}{J.~A. Siegel}, \bibinfo{author}{J.~A. Gilbert},
  \bibinfo{author}{B.~Stephens}, \bibinfo{author}{I.~Asadi},
  \bibinfo{author}{N.~Mahyuddin}, \bibinfo{author}{P.~Shafigh},
  \bibinfo{author}{E.~M. d.~A. Vieira}, \bibinfo{author}{L.~B. da~Silva},
  \bibinfo{author}{E.~L. de~Souza},
\newblock \bibinfo{title}{{A survey of evaluation methods used for holistic
  comfort assessment}},
\newblock \bibinfo{journal}{PLoS ONE} \bibinfo{volume}{953-954}
  (\bibinfo{year}{2016}) \bibinfo{pages}{1513--1519}.
\bibitem[{Tang et~al.(2020)Tang, Ding, and Singer}]{Tang2020}
\bibinfo{author}{H.~Tang}, \bibinfo{author}{Y.~Ding},
  \bibinfo{author}{B.~Singer},
\newblock \bibinfo{title}{{Interactions and comprehensive effect of indoor
  environmental quality factors on occupant satisfaction}},
\newblock \bibinfo{journal}{Building and Environment} \bibinfo{volume}{167}
  (\bibinfo{year}{2020}).
\bibitem[{Henry and Emery(1986)}]{henry1986effect}
\bibinfo{author}{C.~Henry}, \bibinfo{author}{B.~Emery},
\newblock \bibinfo{title}{Effect of spiced food on metabolic rate.},
\newblock \bibinfo{journal}{Human nutrition. Clinical nutrition}
  \bibinfo{volume}{40} (\bibinfo{year}{1986}) \bibinfo{pages}{165--168}.
\bibitem[{Swaminathan et~al.(1985)Swaminathan, King, Holmfield, Siwek, Baker,
  and Wales}]{swaminathan1985thermic}
\bibinfo{author}{R.~Swaminathan}, \bibinfo{author}{R.~King},
  \bibinfo{author}{J.~Holmfield}, \bibinfo{author}{R.~Siwek},
  \bibinfo{author}{M.~Baker}, \bibinfo{author}{J.~Wales},
\newblock \bibinfo{title}{Thermic effect of feeding carbohydrate, fat, protein
  and mixed meal in lean and obese subjects},
\newblock \bibinfo{journal}{The American journal of clinical nutrition}
  \bibinfo{volume}{42} (\bibinfo{year}{1985}) \bibinfo{pages}{177--181}.
\bibitem[{Webster et~al.(2007)Webster, Bauman, and Anwar}]{Webster2007}
\bibinfo{author}{T.~Webster}, \bibinfo{author}{F.~Bauman},
  \bibinfo{author}{G.~Anwar},
\newblock \bibinfo{title}{{CBE Portable Wireless Monitoring System (PWMS): UFAD
  Systems Commissioning Cart Design Specifications and Operating Manual}},
\newblock \bibinfo{journal}{Internal Report, Center for the Built Environment,
  UC Berkeley}  (\bibinfo{year}{2007}) \bibinfo{pages}{4}.
\bibitem[{Choi et~al.(2012)Choi, Loftness, and Aziz}]{Choi2012}
\bibinfo{author}{J.~H. Choi}, \bibinfo{author}{V.~Loftness},
  \bibinfo{author}{A.~Aziz},
\newblock \bibinfo{title}{{Post-occupancy evaluation of 20 office buildings as
  basis for future IEQ standards and guidelines}},
\newblock in: \bibinfo{booktitle}{Energy and Buildings},
  volume~\bibinfo{volume}{46}, pp. \bibinfo{pages}{167--175}.
\bibitem[{Kim(2013)}]{Kim2013}
\bibinfo{author}{H.~Kim},
\newblock \bibinfo{title}{{Methodology for Rating a Building's Overall
  Performance based on the ASHRAE/CIBSE/USGBC Performance Measurement Protocols
  for Commercial Buildings}},
\newblock \bibinfo{journal}{Journal of Chemical Information and Modeling}
  \bibinfo{volume}{53} (\bibinfo{year}{2013}) \bibinfo{pages}{1689--1699}.
\bibitem[{Parkinson et~al.(2019)Parkinson, Parkinson, and
  de~Dear}]{Parkinson2019}
\bibinfo{author}{T.~Parkinson}, \bibinfo{author}{A.~Parkinson},
  \bibinfo{author}{R.~de~Dear},
\newblock \bibinfo{title}{{Continuous IEQ monitoring system: Context and
  development}},
\newblock \bibinfo{journal}{Building and Environment} \bibinfo{volume}{149}
  (\bibinfo{year}{2019}) \bibinfo{pages}{15--25}.
\bibitem[{Nicol and Wilson(2011)}]{Nicol2011}
\bibinfo{author}{J.~F. Nicol}, \bibinfo{author}{M.~Wilson},
\newblock \bibinfo{title}{{A critique of European Standard EN 15251: Strengths,
  weaknesses and lessons for future standards}},
\newblock \bibinfo{journal}{Building Research and Information}
  \bibinfo{volume}{39} (\bibinfo{year}{2011}) \bibinfo{pages}{183--193}.
\bibitem[{Schiavon et~al.(2017)Schiavon, Yang, Donner, Chang, and
  Nazaroff}]{schiavon2017thermal}
\bibinfo{author}{S.~Schiavon}, \bibinfo{author}{B.~Yang},
  \bibinfo{author}{Y.~Donner}, \bibinfo{author}{V.-C. Chang},
  \bibinfo{author}{W.~W. Nazaroff},
\newblock \bibinfo{title}{Thermal comfort, perceived air quality, and cognitive
  performance when personally controlled air movement is used by tropically
  acclimatized persons},
\newblock \bibinfo{journal}{Indoor air} \bibinfo{volume}{27}
  (\bibinfo{year}{2017}) \bibinfo{pages}{690--702}.
\bibitem[{Hodder and Parsons(2007)}]{hodder2007effects}
\bibinfo{author}{S.~G. Hodder}, \bibinfo{author}{K.~Parsons},
\newblock \bibinfo{title}{The effects of solar radiation on thermal comfort},
\newblock \bibinfo{journal}{International journal of biometeorology}
  \bibinfo{volume}{51} (\bibinfo{year}{2007}) \bibinfo{pages}{233--250}.
\bibitem[{Kr{\"a}uchi(2002)}]{krauchi2002circadian}
\bibinfo{author}{K.~Kr{\"a}uchi},
\newblock \bibinfo{title}{How is the circadian rhythm of core body temperature
  regulated?},
\newblock \bibinfo{journal}{Clinical Autonomic Research} \bibinfo{volume}{12}
  (\bibinfo{year}{2002}) \bibinfo{pages}{147--149}.
\bibitem[{Chinazzo et~al.(2019)Chinazzo, Wienold, and
  Andersen}]{chinazzo2019daylight}
\bibinfo{author}{G.~Chinazzo}, \bibinfo{author}{J.~Wienold},
  \bibinfo{author}{M.~Andersen},
\newblock \bibinfo{title}{Daylight affects human thermal perception},
\newblock \bibinfo{journal}{Scientific reports} \bibinfo{volume}{9}
  (\bibinfo{year}{2019}) \bibinfo{pages}{1--15}.
\bibitem[{Halawa et~al.(2014)Halawa, van Hoof, and
  Soebarto}]{halawa2014impacts}
\bibinfo{author}{E.~Halawa}, \bibinfo{author}{J.~van Hoof},
  \bibinfo{author}{V.~Soebarto},
\newblock \bibinfo{title}{The impacts of the thermal radiation field on thermal
  comfort, energy consumption and control—a critical overview},
\newblock \bibinfo{journal}{Renewable and Sustainable Energy Reviews}
  \bibinfo{volume}{37} (\bibinfo{year}{2014}) \bibinfo{pages}{907--918}.
\bibitem[{Fukazawa and Havenith(2009)}]{fukazawa2009differences}
\bibinfo{author}{T.~Fukazawa}, \bibinfo{author}{G.~Havenith},
\newblock \bibinfo{title}{Differences in comfort perception in relation to
  local and whole body skin wettedness},
\newblock \bibinfo{journal}{European journal of applied physiology}
  \bibinfo{volume}{106} (\bibinfo{year}{2009}) \bibinfo{pages}{15--24}.
\bibitem[{Kingma et~al.(2012)Kingma, Frijns, and van
  Marken~Lichtenbelt}]{kingma2012thermoneutral}
\bibinfo{author}{B.~Kingma}, \bibinfo{author}{A.~Frijns},
  \bibinfo{author}{W.~van Marken~Lichtenbelt},
\newblock \bibinfo{title}{The thermoneutral zone: implications for metabolic
  studies},
\newblock \bibinfo{journal}{Front Biosci (Elite Ed)} \bibinfo{volume}{4}
  (\bibinfo{year}{2012}) \bibinfo{pages}{1975--1985}.
\bibitem[{Tikuisis and Ducharme(1996)}]{tikuisis1996effect}
\bibinfo{author}{P.~Tikuisis}, \bibinfo{author}{M.~B. Ducharme},
\newblock \bibinfo{title}{The effect of postural changes on body temperatures
  and heat balance},
\newblock \bibinfo{journal}{European journal of applied physiology and
  occupational physiology} \bibinfo{volume}{72} (\bibinfo{year}{1996})
  \bibinfo{pages}{451--459}.
\bibitem[{Gagge et~al.(1967)Gagge, Stolwijk, and Hardy}]{gagge1967comfort}
\bibinfo{author}{A.~P. Gagge}, \bibinfo{author}{J.~Stolwijk},
  \bibinfo{author}{J.~Hardy},
\newblock \bibinfo{title}{Comfort and thermal sensations and associated
  physiological responses at various ambient temperatures},
\newblock \bibinfo{journal}{Environmental research} \bibinfo{volume}{1}
  (\bibinfo{year}{1967}) \bibinfo{pages}{1--20}.
\bibitem[{Gaesser and Brooks(1975)}]{gaesser1975muscular}
\bibinfo{author}{G.~A. Gaesser}, \bibinfo{author}{G.~A. Brooks},
\newblock \bibinfo{title}{Muscular efficiency during steady-rate exercise:
  effects of speed and work rate},
\newblock \bibinfo{journal}{Journal of applied physiology} \bibinfo{volume}{38}
  (\bibinfo{year}{1975}) \bibinfo{pages}{1132--1139}.
\bibitem[{Havenith et~al.(2015)Havenith, Kuklane, Fan, Hodder, Ouzzahra,
  Lundgren, Au, and Loveday}]{havenith2015database}
\bibinfo{author}{G.~Havenith}, \bibinfo{author}{K.~Kuklane},
  \bibinfo{author}{J.~Fan}, \bibinfo{author}{S.~Hodder},
  \bibinfo{author}{Y.~Ouzzahra}, \bibinfo{author}{K.~Lundgren},
  \bibinfo{author}{Y.~Au}, \bibinfo{author}{D.~L. Loveday},
\newblock \bibinfo{title}{A database of static clothing thermal insulation and
  vapor permeability values of non-western ensembles for use in ashrae standard
  55, iso 7730, and iso 9920 ch-15-018 (rp-1504)}  (\bibinfo{year}{2015}).
\bibitem[{Johnson and Kellogg~Jr(2010)}]{johnson2010thermoregulatory}
\bibinfo{author}{J.~M. Johnson}, \bibinfo{author}{D.~L. Kellogg~Jr},
\newblock \bibinfo{title}{Thermoregulatory and thermal control in the human
  cutaneous circulation},
\newblock \bibinfo{journal}{Front Biosci (Schol Ed)} \bibinfo{volume}{2}
  (\bibinfo{year}{2010}) \bibinfo{pages}{825--853}.
\bibitem[{Zhang and Hedge(2014)}]{zhang2014laptop}
\bibinfo{author}{H.~Zhang}, \bibinfo{author}{A.~Hedge},
\newblock \bibinfo{title}{Laptop heat and models of user thermal discomfort},
\newblock in: \bibinfo{booktitle}{Proceedings of the Human Factors and
  Ergonomics Society Annual Meeting}, volume~\bibinfo{volume}{58},
  \bibinfo{organization}{SAGE Publications Sage CA: Los Angeles, CA}, pp.
  \bibinfo{pages}{1456--1460}.
\bibitem[{Karjalainen(2012)}]{karjalainen2012thermal}
\bibinfo{author}{S.~Karjalainen},
\newblock \bibinfo{title}{Thermal comfort and gender: a literature review},
\newblock \bibinfo{journal}{Indoor air} \bibinfo{volume}{22}
  (\bibinfo{year}{2012}) \bibinfo{pages}{96--109}.
\bibitem[{Huang et~al.(2010)Huang, Wang, and Lin}]{huang2010influence}
\bibinfo{author}{H.-W. Huang}, \bibinfo{author}{W.-C. Wang},
  \bibinfo{author}{C.-C.~K. Lin},
\newblock \bibinfo{title}{Influence of age on thermal thresholds, thermal pain
  thresholds, and reaction time},
\newblock \bibinfo{journal}{Journal of Clinical Neuroscience}
  \bibinfo{volume}{17} (\bibinfo{year}{2010}) \bibinfo{pages}{722--726}.
\bibitem[{Havenith et~al.(2002)Havenith, Holm{\'e}r, and
  Parsons}]{havenith2002personal}
\bibinfo{author}{G.~Havenith}, \bibinfo{author}{I.~Holm{\'e}r},
  \bibinfo{author}{K.~Parsons},
\newblock \bibinfo{title}{Personal factors in thermal comfort assessment:
  clothing properties and metabolic heat production},
\newblock \bibinfo{journal}{Energy and buildings} \bibinfo{volume}{34}
  (\bibinfo{year}{2002}) \bibinfo{pages}{581--591}.
\bibitem[{Frank et~al.(1999)Frank, Raja, Bulcao, and
  Goldstein}]{frank1999relative}
\bibinfo{author}{S.~M. Frank}, \bibinfo{author}{S.~N. Raja},
  \bibinfo{author}{C.~F. Bulcao}, \bibinfo{author}{D.~S. Goldstein},
\newblock \bibinfo{title}{Relative contribution of core and cutaneous
  temperatures to thermal comfort and autonomic responses in humans},
\newblock \bibinfo{journal}{Journal of Applied physiology} \bibinfo{volume}{86}
  (\bibinfo{year}{1999}) \bibinfo{pages}{1588--1593}.
\bibitem[{Boulant(2000)}]{boulant2000role}
\bibinfo{author}{J.~A. Boulant},
\newblock \bibinfo{title}{Role of the preoptic-anterior hypothalamus in
  thermoregulation and fever},
\newblock \bibinfo{journal}{Clinical infectious diseases} \bibinfo{volume}{31}
  (\bibinfo{year}{2000}) \bibinfo{pages}{S157--S161}.
\bibitem[{Cabanac(1971)}]{cabanac1971physiological}
\bibinfo{author}{M.~Cabanac},
\newblock \bibinfo{title}{Physiological role of pleasure},
\newblock \bibinfo{journal}{Science} \bibinfo{volume}{173}
  (\bibinfo{year}{1971}) \bibinfo{pages}{1103--1107}.
\bibitem[{Brainard et~al.(2001)Brainard, Hanifin, Greeson, Byrne, Glickman,
  Gerner, and Rollag}]{brainard2001action}
\bibinfo{author}{G.~C. Brainard}, \bibinfo{author}{J.~P. Hanifin},
  \bibinfo{author}{J.~M. Greeson}, \bibinfo{author}{B.~Byrne},
  \bibinfo{author}{G.~Glickman}, \bibinfo{author}{E.~Gerner},
  \bibinfo{author}{M.~D. Rollag},
\newblock \bibinfo{title}{Action spectrum for melatonin regulation in humans:
  evidence for a novel circadian photoreceptor},
\newblock \bibinfo{journal}{Journal of Neuroscience} \bibinfo{volume}{21}
  (\bibinfo{year}{2001}) \bibinfo{pages}{6405--6412}.
\bibitem[{Perez et~al.(1990)Perez, Ineichen, Seals, Michalsky, and
  Stewart}]{perez1990modeling}
\bibinfo{author}{R.~Perez}, \bibinfo{author}{P.~Ineichen},
  \bibinfo{author}{R.~Seals}, \bibinfo{author}{J.~Michalsky},
  \bibinfo{author}{R.~Stewart},
\newblock \bibinfo{title}{Modeling daylight availability and irradiance
  components from direct and global irradiance},
\newblock \bibinfo{journal}{Solar energy} \bibinfo{volume}{44}
  (\bibinfo{year}{1990}) \bibinfo{pages}{271--289}.
\bibitem[{Leather et~al.(1998)Leather, Pyrgas, Beale, and
  Lawrence}]{leather1998windows}
\bibinfo{author}{P.~Leather}, \bibinfo{author}{M.~Pyrgas},
  \bibinfo{author}{D.~Beale}, \bibinfo{author}{C.~Lawrence},
\newblock \bibinfo{title}{Windows in the workplace: Sunlight, view, and
  occupational stress},
\newblock \bibinfo{journal}{Environment and behavior} \bibinfo{volume}{30}
  (\bibinfo{year}{1998}) \bibinfo{pages}{739--762}.
\bibitem[{Dai et~al.(2016)Dai, Hao, Lin, and Cui}]{dai2016spectral}
\bibinfo{author}{Q.~Dai}, \bibinfo{author}{L.~Hao}, \bibinfo{author}{Y.~Lin},
  \bibinfo{author}{Z.~Cui},
\newblock \bibinfo{title}{Spectral optimization simulation of white light based
  on the photopic eye-sensitivity curve},
\newblock \bibinfo{journal}{Journal of Applied Physics} \bibinfo{volume}{119}
  (\bibinfo{year}{2016}) \bibinfo{pages}{053103}.
\bibitem[{Slater and Boyce(1990)}]{slater1990illuminance}
\bibinfo{author}{A.~I. Slater}, \bibinfo{author}{P.~R. Boyce},
\newblock \bibinfo{title}{Illuminance uniformity on desks: Where is the
  limit?},
\newblock \bibinfo{journal}{Lighting research \& technology}
  \bibinfo{volume}{22} (\bibinfo{year}{1990}) \bibinfo{pages}{165--174}.
\bibitem[{Baron et~al.(1992)Baron, Rea, and Daniels}]{baron1992effects}
\bibinfo{author}{R.~A. Baron}, \bibinfo{author}{M.~S. Rea},
  \bibinfo{author}{S.~G. Daniels},
\newblock \bibinfo{title}{Effects of indoor lighting (illuminance and spectral
  distribution) on the performance of cognitive tasks and interpersonal
  behaviors: The potential mediating role of positive affect},
\newblock \bibinfo{journal}{Motivation and emotion} \bibinfo{volume}{16}
  (\bibinfo{year}{1992}) \bibinfo{pages}{1--33}.
\bibitem[{Wilkins et~al.(1989)Wilkins, Nimmo-Smith, Slater, and
  Bedocs}]{wilkins1989fluorescent}
\bibinfo{author}{A.~J. Wilkins}, \bibinfo{author}{I.~Nimmo-Smith},
  \bibinfo{author}{A.~I. Slater}, \bibinfo{author}{L.~Bedocs},
\newblock \bibinfo{title}{Fluorescent lighting, headaches and eyestrain},
\newblock \bibinfo{journal}{Lighting Research \& Technology}
  \bibinfo{volume}{21} (\bibinfo{year}{1989}) \bibinfo{pages}{11--18}.
\bibitem[{Main et~al.(1997)Main, Dowson, and Gross}]{main1997photophobia}
\bibinfo{author}{A.~Main}, \bibinfo{author}{A.~Dowson},
  \bibinfo{author}{M.~Gross},
\newblock \bibinfo{title}{Photophobia and phonophobia in migraineurs between
  attacks},
\newblock \bibinfo{journal}{Headache: The Journal of Head and Face Pain}
  \bibinfo{volume}{37} (\bibinfo{year}{1997}) \bibinfo{pages}{492--495}.
\bibitem[{Yin et~al.(2018)Yin, Zhu, MacNaughton, Allen, and
  Spengler}]{yin2018physiological}
\bibinfo{author}{J.~Yin}, \bibinfo{author}{S.~Zhu},
  \bibinfo{author}{P.~MacNaughton}, \bibinfo{author}{J.~G. Allen},
  \bibinfo{author}{J.~D. Spengler},
\newblock \bibinfo{title}{Physiological and cognitive performance of exposure
  to biophilic indoor environment},
\newblock \bibinfo{journal}{Building and Environment} \bibinfo{volume}{132}
  (\bibinfo{year}{2018}) \bibinfo{pages}{255--262}.
\bibitem[{Hopkinson(1972)}]{hopkinson1972glare}
\bibinfo{author}{R.~G. Hopkinson},
\newblock \bibinfo{title}{Glare from daylighting in buildings},
\newblock \bibinfo{journal}{Applied ergonomics} \bibinfo{volume}{3}
  (\bibinfo{year}{1972}) \bibinfo{pages}{206--215}.
\bibitem[{Pierrette et~al.(2015)Pierrette, Parizet, Chevret, and
  Chatillon}]{pierrette2015noise}
\bibinfo{author}{M.~Pierrette}, \bibinfo{author}{E.~Parizet},
  \bibinfo{author}{P.~Chevret}, \bibinfo{author}{J.~Chatillon},
\newblock \bibinfo{title}{Noise effect on comfort in open-space offices:
  development of an assessment questionnaire},
\newblock \bibinfo{journal}{Ergonomics} \bibinfo{volume}{58}
  (\bibinfo{year}{2015}) \bibinfo{pages}{96--106}.
\bibitem[{Job(1988)}]{job1988community}
\bibinfo{author}{R.~Job},
\newblock \bibinfo{title}{Community response to noise: A review of factors
  influencing the relationship between noise exposure and reaction},
\newblock \bibinfo{journal}{The Journal of the Acoustical Society of America}
  \bibinfo{volume}{83} (\bibinfo{year}{1988}) \bibinfo{pages}{991--1001}.
\bibitem[{Kim and De~Dear(2013)}]{kim2013workspace}
\bibinfo{author}{J.~Kim}, \bibinfo{author}{R.~De~Dear},
\newblock \bibinfo{title}{Workspace satisfaction: The privacy-communication
  trade-off in open-plan offices},
\newblock \bibinfo{journal}{Journal of Environmental Psychology}
  \bibinfo{volume}{36} (\bibinfo{year}{2013}) \bibinfo{pages}{18--26}.
\bibitem[{Templeton and Saunders(2014)}]{templeton2014acoustic}
\bibinfo{author}{D.~Templeton}, \bibinfo{author}{D.~Saunders},
  \bibinfo{title}{Acoustic design}, \bibinfo{publisher}{Elsevier},
  \bibinfo{year}{2014}.
\bibitem[{Lee and Brand(2010)}]{lee2010can}
\bibinfo{author}{S.~Y. Lee}, \bibinfo{author}{J.~Brand},
\newblock \bibinfo{title}{Can personal control over the physical environment
  ease distractions in office workplaces?},
\newblock \bibinfo{journal}{Ergonomics} \bibinfo{volume}{53}
  (\bibinfo{year}{2010}) \bibinfo{pages}{324--335}.
\bibitem[{Kjellberg and Sk{\"o}ldstr{\"o}m(1991)}]{kjellberg1991noise}
\bibinfo{author}{A.~Kjellberg}, \bibinfo{author}{B.~Sk{\"o}ldstr{\"o}m},
\newblock \bibinfo{title}{Noise annoyance during the performance of different
  nonauditory tasks},
\newblock \bibinfo{journal}{Perceptual and motor skills} \bibinfo{volume}{73}
  (\bibinfo{year}{1991}) \bibinfo{pages}{39--49}.
\bibitem[{Banbury and Berry(2005)}]{banbury2005office}
\bibinfo{author}{S.~P. Banbury}, \bibinfo{author}{D.~C. Berry},
\newblock \bibinfo{title}{Office noise and employee concentration: Identifying
  causes of disruption and potential improvements},
\newblock \bibinfo{journal}{Ergonomics} \bibinfo{volume}{48}
  (\bibinfo{year}{2005}) \bibinfo{pages}{25--37}.
\bibitem[{Cheung et~al.(2019)Cheung, Schiavon, Parkinson, Li, and
  Brager}]{cheung2019analysis}
\bibinfo{author}{T.~Cheung}, \bibinfo{author}{S.~Schiavon},
  \bibinfo{author}{T.~Parkinson}, \bibinfo{author}{P.~Li},
  \bibinfo{author}{G.~Brager},
\newblock \bibinfo{title}{Analysis of the accuracy on pmv--ppd model using the
  ashrae global thermal comfort database ii},
\newblock \bibinfo{journal}{Building and Environment}  (\bibinfo{year}{2019}).
\bibitem[{Stone et~al.(2007)Stone, Shiffman, Atienza, Nebeling, Stone,
  Shiffman, Atienza, and Nebeling}]{stone2007historical}
\bibinfo{author}{A.~A. Stone}, \bibinfo{author}{S.~Shiffman},
  \bibinfo{author}{A.~A. Atienza}, \bibinfo{author}{L.~Nebeling},
  \bibinfo{author}{A.~Stone}, \bibinfo{author}{S.~Shiffman},
  \bibinfo{author}{A.~Atienza}, \bibinfo{author}{L.~Nebeling},
\newblock \bibinfo{title}{Historical roots and rationale of ecological
  momentary assessment (ema)},
\newblock \bibinfo{journal}{The science of real-time data capture: Self-reports
  in health research}  (\bibinfo{year}{2007}) \bibinfo{pages}{3--10}.
\bibitem[{Intille et~al.(2016)Intille, Haynes, Maniar, Ponnada, and
  Manjourides}]{intille2016muema}
\bibinfo{author}{S.~Intille}, \bibinfo{author}{C.~Haynes},
  \bibinfo{author}{D.~Maniar}, \bibinfo{author}{A.~Ponnada},
  \bibinfo{author}{J.~Manjourides},
\newblock \bibinfo{title}{$\mu$ema: Microinteraction-based ecological momentary
  assessment (ema) using a smartwatch},
\newblock in: \bibinfo{booktitle}{Proceedings of the 2016 ACM International
  Joint Conference on Pervasive and Ubiquitous Computing},
  \bibinfo{organization}{ACM}, pp. \bibinfo{pages}{1124--1128}.
\bibitem[{Wang et~al.(2014)Wang, Amin, Li, Abdelzaher, Kaplan, Gu, Pan, Liu,
  Aggarwal, Ganti, Wang, Mohapatra, Szymanski, and Le}]{Wang2014-qj}
\bibinfo{author}{D.~Wang}, \bibinfo{author}{M.~T. Amin},
  \bibinfo{author}{S.~Li}, \bibinfo{author}{T.~Abdelzaher},
  \bibinfo{author}{L.~Kaplan}, \bibinfo{author}{S.~Gu},
  \bibinfo{author}{C.~Pan}, \bibinfo{author}{H.~Liu}, \bibinfo{author}{C.~C.
  Aggarwal}, \bibinfo{author}{R.~Ganti}, \bibinfo{author}{X.~Wang},
  \bibinfo{author}{P.~Mohapatra}, \bibinfo{author}{B.~Szymanski},
  \bibinfo{author}{H.~Le},
\newblock \bibinfo{title}{Using humans as sensors: an estimation-theoretic
  perspective},
\newblock in: \bibinfo{booktitle}{Proceedings of the 13th international
  symposium on Information processing in sensor networks}, IPSN '14,
  \bibinfo{publisher}{IEEE Press}, \bibinfo{year}{2014}, pp.
  \bibinfo{pages}{35--46}.
\bibitem[{Avvenuti et~al.(2016)Avvenuti, Cimino, Cresci, Marchetti, and
  Tesconi}]{Avvenuti2016-qy}
\bibinfo{author}{M.~Avvenuti}, \bibinfo{author}{M.~G. C.~A. Cimino},
  \bibinfo{author}{S.~Cresci}, \bibinfo{author}{A.~Marchetti},
  \bibinfo{author}{M.~Tesconi},
\newblock \bibinfo{title}{A framework for detecting unfolding emergencies using
  humans as sensors},
\newblock \bibinfo{journal}{Springerplus} \bibinfo{volume}{5}
  (\bibinfo{year}{2016}) \bibinfo{pages}{43}.
\bibitem[{Vielberth et~al.(2019)Vielberth, Menges, and
  Pernul}]{Vielberth2019-nl}
\bibinfo{author}{M.~Vielberth}, \bibinfo{author}{F.~Menges},
  \bibinfo{author}{G.~Pernul},
\newblock \bibinfo{title}{Human-as-a-security-sensor for harvesting threat
  intelligence},
\newblock \bibinfo{journal}{Cybersecurity} \bibinfo{volume}{2}
  (\bibinfo{year}{2019}) \bibinfo{pages}{23}.
\bibitem[{Heinzerling et~al.(2013)Heinzerling, Schiavon, Webster, and
  Arens}]{Heinzerling2013}
\bibinfo{author}{D.~Heinzerling}, \bibinfo{author}{S.~Schiavon},
  \bibinfo{author}{T.~Webster}, \bibinfo{author}{E.~Arens},
\newblock \bibinfo{title}{{Indoor environmental quality assessment models: A
  literature review and a proposed weighting and classification scheme}}
  \bibinfo{volume}{70} (\bibinfo{year}{2013}) \bibinfo{pages}{210--222}.
\bibitem[{Ncube and Riffat(2012)}]{Ncube2012}
\bibinfo{author}{M.~Ncube}, \bibinfo{author}{S.~Riffat},
\newblock \bibinfo{title}{{Developing an indoor environment quality tool for
  assessment of mechanically ventilated office buildings in the UK - A
  preliminary study}},
\newblock \bibinfo{journal}{Building and Environment} \bibinfo{volume}{53}
  (\bibinfo{year}{2012}) \bibinfo{pages}{26--33}.
\bibitem[{Wong et~al.(2008)Wong, Mui, and Hui}]{Wong2008}
\bibinfo{author}{L.~T. Wong}, \bibinfo{author}{K.~W. Mui},
  \bibinfo{author}{P.~S. Hui},
\newblock \bibinfo{title}{{A multivariate-logistic model for acceptance of
  indoor environmental quality (IEQ) in offices}},
\newblock \bibinfo{journal}{Building and Environment} \bibinfo{volume}{43}
  (\bibinfo{year}{2008}) \bibinfo{pages}{1--6}.
\bibitem[{Lai et~al.(2009)Lai, Mui, Wong, and Law}]{Lai2009}
\bibinfo{author}{A.~C. Lai}, \bibinfo{author}{K.~W. Mui},
  \bibinfo{author}{L.~T. Wong}, \bibinfo{author}{L.~Y. Law},
\newblock \bibinfo{title}{{An evaluation model for indoor environmental quality
  (IEQ) acceptance in residential buildings}},
\newblock \bibinfo{journal}{Energy and Buildings} \bibinfo{volume}{41}
  (\bibinfo{year}{2009}) \bibinfo{pages}{930--936}.
\bibitem[{Cohen et~al.(2001)Cohen, Standeven, Bordass, and Leaman}]{Cohen2001}
\bibinfo{author}{R.~Cohen}, \bibinfo{author}{M.~Standeven},
  \bibinfo{author}{B.~Bordass}, \bibinfo{author}{A.~Leaman},
\newblock \bibinfo{title}{{Assessing building performance in use 1: The Probe
  process}},
\newblock \bibinfo{journal}{Building Research and Information}
  \bibinfo{volume}{29} (\bibinfo{year}{2001}) \bibinfo{pages}{85--102}.
\bibitem[{Webster et~al.(2007)Webster, Arens, Anwar, Bonnell, Bauman, and
  Brown}]{webster2007ufad}
\bibinfo{author}{T.~Webster}, \bibinfo{author}{E.~Arens},
  \bibinfo{author}{G.~Anwar}, \bibinfo{author}{J.~Bonnell},
  \bibinfo{author}{F.~Bauman}, \bibinfo{author}{C.~Brown},
\newblock \bibinfo{title}{Ufad commissioning cart: Design specifications and
  operating manual}  (\bibinfo{year}{2007}).
\bibitem[{Jin et~al.(2018)Jin, Liu, Schiavon, and Spanos}]{jin2018automated}
\bibinfo{author}{M.~Jin}, \bibinfo{author}{S.~Liu},
  \bibinfo{author}{S.~Schiavon}, \bibinfo{author}{C.~Spanos},
\newblock \bibinfo{title}{Automated mobile sensing: Towards high-granularity
  agile indoor environmental quality monitoring},
\newblock \bibinfo{journal}{Building and Environment} \bibinfo{volume}{127}
  (\bibinfo{year}{2018}) \bibinfo{pages}{268--276}.
\bibitem[{Porter et~al.(2004)Porter, Whitcomb, and
  Weitzer}]{porter2004multiple}
\bibinfo{author}{S.~R. Porter}, \bibinfo{author}{M.~E. Whitcomb},
  \bibinfo{author}{W.~H. Weitzer},
\newblock \bibinfo{title}{Multiple surveys of students and survey fatigue},
\newblock \bibinfo{journal}{New Directions for Institutional Research}
  \bibinfo{volume}{2004} (\bibinfo{year}{2004}) \bibinfo{pages}{63--73}.
\bibitem[{Liu et~al.(2019)Liu, Schiavon, Das, Jin, and Spanos}]{Liu2019-pi}
\bibinfo{author}{S.~Liu}, \bibinfo{author}{S.~Schiavon}, \bibinfo{author}{H.~P.
  Das}, \bibinfo{author}{M.~Jin}, \bibinfo{author}{C.~J. Spanos},
\newblock \bibinfo{title}{Personal thermal comfort models with wearable
  sensors},
\newblock \bibinfo{journal}{Build. Environ.} \bibinfo{volume}{162}
  (\bibinfo{year}{2019}) \bibinfo{pages}{106281}.
\bibitem[{Aryal and Becerik-Gerber(2019)}]{Aryal2019-bx}
\bibinfo{author}{A.~Aryal}, \bibinfo{author}{B.~Becerik-Gerber},
\newblock \bibinfo{title}{A comparative study of predicting individual thermal
  sensation and satisfaction using wrist-worn temperature sensor, thermal
  camera and ambient temperature sensor},
\newblock \bibinfo{journal}{Build. Environ.} \bibinfo{volume}{160}
  (\bibinfo{year}{2019}) \bibinfo{pages}{106223}.
\bibitem[{Clear et~al.(2018)Clear, {Mitchell Finnigan}, Olivier, and
  Comber}]{Clear2018}
\bibinfo{author}{A.~K. Clear}, \bibinfo{author}{S.~{Mitchell Finnigan}},
  \bibinfo{author}{P.~Olivier}, \bibinfo{author}{R.~Comber},
\newblock \bibinfo{title}{{ThermoKiosk: Investigating Roles for Digital Surveys
  of Thermal Experience in Workplace Comfort Management}},
\newblock \bibinfo{journal}{Proc. of CHI}  (\bibinfo{year}{2018})
  \bibinfo{pages}{1--12}.
\bibitem[{Engelen and Held(2019)}]{engelen2019understanding}
\bibinfo{author}{L.~Engelen}, \bibinfo{author}{F.~Held},
\newblock \bibinfo{title}{Understanding the office: Using ecological momentary
  assessment to measure activities, posture, social interactions, mood, and
  work performance at the workplace},
\newblock \bibinfo{journal}{Buildings} \bibinfo{volume}{9}
  (\bibinfo{year}{2019}) \bibinfo{pages}{54}.
\bibitem[{Monnot et~al.(2016)Monnot, Wilhelm, Piliouras, Zhou, Dahlmeier, Lu,
  and Jin}]{Monnot2016-fl}
\bibinfo{author}{B.~Monnot}, \bibinfo{author}{E.~Wilhelm},
  \bibinfo{author}{G.~Piliouras}, \bibinfo{author}{Y.~Zhou},
  \bibinfo{author}{D.~Dahlmeier}, \bibinfo{author}{H.~Y. Lu},
  \bibinfo{author}{W.~Jin},
\newblock \bibinfo{title}{Inferring activities and optimal trips: Lessons from
  singapore's national science experiment},
\newblock in: \bibinfo{booktitle}{Complex Systems Design \& Management Asia},
  \bibinfo{publisher}{Springer International Publishing}, \bibinfo{year}{2016},
  pp. \bibinfo{pages}{247--264}.
\bibitem[{Wilhelm et~al.(2016)Wilhelm, Zhou, Zhang, Kee, Loh, and
  Tippenhauer}]{Wilhelm2016-il}
\bibinfo{author}{E.~Wilhelm}, \bibinfo{author}{Y.~Zhou},
  \bibinfo{author}{N.~Zhang}, \bibinfo{author}{J.~Kee},
  \bibinfo{author}{G.~Loh}, \bibinfo{author}{N.~Tippenhauer},
  \bibinfo{title}{Sensg: Large-scale deployment of wearable sensors for trip
  and transport mode logging}, \bibinfo{type}{Technical Report},
  \bibinfo{year}{2016}.
\bibitem[{Benita et~al.(2019)Benita, Bansal, and Tun{\c c}er}]{Benita2019-dl}
\bibinfo{author}{F.~Benita}, \bibinfo{author}{G.~Bansal},
  \bibinfo{author}{B.~Tun{\c c}er},
\newblock \bibinfo{title}{Public spaces and happiness: Evidence from a
  large-scale field experiment},
\newblock \bibinfo{journal}{Health Place} \bibinfo{volume}{56}
  (\bibinfo{year}{2019}) \bibinfo{pages}{9--18}.
\bibitem[{Ojha et~al.(2019)Ojha, Griego, Kuliga, Bielik, Bu{\v s}, Schaeben,
  Treyer, Standfest, Schneider, K{\"o}nig, Donath, and Schmitt}]{Ojha2019-jx}
\bibinfo{author}{V.~K. Ojha}, \bibinfo{author}{D.~Griego},
  \bibinfo{author}{S.~Kuliga}, \bibinfo{author}{M.~Bielik},
  \bibinfo{author}{P.~Bu{\v s}}, \bibinfo{author}{C.~Schaeben},
  \bibinfo{author}{L.~Treyer}, \bibinfo{author}{M.~Standfest},
  \bibinfo{author}{S.~Schneider}, \bibinfo{author}{R.~K{\"o}nig},
  \bibinfo{author}{D.~Donath}, \bibinfo{author}{G.~Schmitt},
\newblock \bibinfo{title}{Machine learning approaches to understand the
  influence of urban environments on human's physiological response},
\newblock \bibinfo{journal}{Inf. Sci.} \bibinfo{volume}{474}
  (\bibinfo{year}{2019}) \bibinfo{pages}{154--169}.
\bibitem[{Rahaman et~al.(2020)Rahaman, Liono, Ren, Chan, Kudo, Rawling, and
  Salim}]{Rahaman2020-gt}
\bibinfo{author}{M.~S. Rahaman}, \bibinfo{author}{J.~Liono},
  \bibinfo{author}{Y.~Ren}, \bibinfo{author}{J.~Chan},
  \bibinfo{author}{S.~Kudo}, \bibinfo{author}{T.~Rawling},
  \bibinfo{author}{F.~D. Salim},
\newblock \bibinfo{title}{An {Ambient-Physical} system to infer concentration
  in open-plan workplace},
\newblock \bibinfo{journal}{IEEE Internet of Things Journal}
  (\bibinfo{year}{2020}) \bibinfo{pages}{1--1}.
\bibitem[{Sood et~al.(2020)Sood, Janssen, and Miller}]{Sood2020-vz}
\bibinfo{author}{T.~Sood}, \bibinfo{author}{P.~Janssen},
  \bibinfo{author}{C.~Miller},
\newblock \bibinfo{title}{Spacematch: Using environmental preferences to match
  occupants to suitable activity-based workspaces},
\newblock \bibinfo{journal}{Frontiers in Built Environment} \bibinfo{volume}{6}
  (\bibinfo{year}{2020}) \bibinfo{pages}{113}.
\bibitem[{Sood et~al.(2019)Sood, Quintana, Jayathissa, AbdelRahman, and
  Miller}]{Sood2019-af}
\bibinfo{author}{T.~Sood}, \bibinfo{author}{M.~Quintana},
  \bibinfo{author}{P.~Jayathissa}, \bibinfo{author}{M.~AbdelRahman},
  \bibinfo{author}{C.~Miller},
\newblock \bibinfo{title}{The {SDE4} learning trail: Crowdsourcing occupant
  comfort feedback at a net-zero energy building},
\newblock \bibinfo{journal}{J. Phys. Conf. Ser.} \bibinfo{volume}{1343}
  (\bibinfo{year}{2019}) \bibinfo{pages}{012141}.
\bibitem[{Jayathissa et~al.(2019)Jayathissa, Quintana, Sood, Narzarian, and
  Miller}]{jayathissayour}
\bibinfo{author}{P.~Jayathissa}, \bibinfo{author}{M.~Quintana},
  \bibinfo{author}{T.~Sood}, \bibinfo{author}{N.~Narzarian},
  \bibinfo{author}{C.~Miller},
\newblock \bibinfo{title}{Is your clock-face cozie? a smartwatch methodology
  for the in-situ collection of occupant comfort data},
\newblock \bibinfo{journal}{CISBAT - Climate Resilient Cities - Energy
  Efficiency and Renewables in the Digital Era}  (\bibinfo{year}{2019}).
\bibitem[{Abdelrahman et~al.(2019)Abdelrahman, Jayathissa, and
  Miller}]{Abdelrahman2019}
\bibinfo{author}{M.~M. Abdelrahman}, \bibinfo{author}{P.~Jayathissa},
  \bibinfo{author}{C.~Miller},
\newblock \bibinfo{title}{{YAK: An Indoor Positioning App for Spatial-Temporal
  Indoor Environmental Quality Research}}  (\bibinfo{year}{2019}).
\bibitem[{Kim et~al.(2018)Kim, Zhou, Schiavon, Raftery, and
  Brager}]{kim2018paradigm}
\bibinfo{author}{J.~Kim}, \bibinfo{author}{Y.~Zhou},
  \bibinfo{author}{S.~Schiavon}, \bibinfo{author}{P.~Raftery},
  \bibinfo{author}{G.~Brager},
\newblock \bibinfo{title}{Personal comfort models: predicting individuals'
  thermal preference using occupant heating and cooling behavior and machine
  learning},
\newblock \bibinfo{journal}{Building and Environment} \bibinfo{volume}{129}
  (\bibinfo{year}{2018}) \bibinfo{pages}{96--106}.
\bibitem[{Luo et~al.(2020)Luo, Xie, Yan, Ke, Yu, Wang, and Zhang}]{Luo2020}
\bibinfo{author}{M.~Luo}, \bibinfo{author}{J.~Xie}, \bibinfo{author}{Y.~Yan},
  \bibinfo{author}{Z.~Ke}, \bibinfo{author}{P.~Yu}, \bibinfo{author}{Z.~Wang},
  \bibinfo{author}{J.~Zhang},
\newblock \bibinfo{title}{{Comparing machine learning algorithms in predicting
  thermal sensation using ASHRAE Comfort Database II}},
\newblock \bibinfo{journal}{Energy and Buildings} \bibinfo{volume}{210}
  (\bibinfo{year}{2020}) \bibinfo{pages}{109776}.
\bibitem[{Kim et~al.(2018)Kim, Schiavon, and Brager}]{kim2018personal}
\bibinfo{author}{J.~Kim}, \bibinfo{author}{S.~Schiavon},
  \bibinfo{author}{G.~Brager},
\newblock \bibinfo{title}{Personal comfort models--a new paradigm in thermal
  comfort for occupant-centric environmental control},
\newblock \bibinfo{journal}{Building and Environment} \bibinfo{volume}{132}
  (\bibinfo{year}{2018}) \bibinfo{pages}{114--124}.
\bibitem[{Enescu(2017)}]{Enescu2017}
\bibinfo{author}{D.~Enescu},
\newblock \bibinfo{title}{{A review of thermal comfort models and indicators
  for indoor environments}},
\newblock \bibinfo{journal}{Renewable and Sustainable Energy Reviews}
  \bibinfo{volume}{79} (\bibinfo{year}{2017}) \bibinfo{pages}{1353--1379}.
\bibitem[{Barrios and Kleiminger(2017)}]{Barrios2017}
\bibinfo{author}{L.~Barrios}, \bibinfo{author}{W.~Kleiminger},
\newblock \bibinfo{title}{{The Comfstat - Automatically sensing thermal comfort
  for smart thermostats}},
\newblock \bibinfo{journal}{2017 IEEE International Conference on Pervasive
  Computing and Communications, PerCom 2017}  (\bibinfo{year}{2017})
  \bibinfo{pages}{257--266}.
\bibitem[{Park and Nagy(2018)}]{Park2018}
\bibinfo{author}{J.~Y. Park}, \bibinfo{author}{Z.~Nagy},
\newblock \bibinfo{title}{{Comprehensive analysis of the relationship between
  thermal comfort and building control research - A data-driven literature
  review}},
\newblock \bibinfo{journal}{Renewable and Sustainable Energy Reviews}
  \bibinfo{volume}{82} (\bibinfo{year}{2018}) \bibinfo{pages}{2664--2679}.
\bibitem[{Zhang et~al.(2015)Zhang, Arens, and Zhai}]{Zhang2015}
\bibinfo{author}{H.~Zhang}, \bibinfo{author}{E.~Arens},
  \bibinfo{author}{Y.~Zhai},
\newblock \bibinfo{title}{{A review of the corrective power of personal comfort
  systems in non-neutral ambient environments}},
\newblock \bibinfo{journal}{Building and Environment} \bibinfo{volume}{91}
  (\bibinfo{year}{2015}) \bibinfo{pages}{15--41}.
\bibitem[{Park et~al.(2019{\natexlab{a}})Park, Ouf, Gunay, Peng, O'Brien,
  Kj{\ae}rgaard, and Nagy}]{Park2019-nm}
\bibinfo{author}{J.~Y. Park}, \bibinfo{author}{M.~M. Ouf},
  \bibinfo{author}{B.~Gunay}, \bibinfo{author}{Y.~Peng},
  \bibinfo{author}{W.~O'Brien}, \bibinfo{author}{M.~B. Kj{\ae}rgaard},
  \bibinfo{author}{Z.~Nagy},
\newblock \bibinfo{title}{A critical review of field implementations of
  occupant-centric building controls},
\newblock \bibinfo{journal}{Build. Environ.} \bibinfo{volume}{165}
  (\bibinfo{year}{2019}{\natexlab{a}}) \bibinfo{pages}{106351}.
\bibitem[{Park et~al.(2019{\natexlab{b}})Park, Dougherty, Fritz, and
  Nagy}]{Park2019-qm}
\bibinfo{author}{J.~Y. Park}, \bibinfo{author}{T.~Dougherty},
  \bibinfo{author}{H.~Fritz}, \bibinfo{author}{Z.~Nagy},
\newblock \bibinfo{title}{{LightLearn}: An adaptive and occupant centered
  controller for lighting based on reinforcement learning},
\newblock \bibinfo{journal}{Build. Environ.} \bibinfo{volume}{147}
  (\bibinfo{year}{2019}{\natexlab{b}}) \bibinfo{pages}{397--414}.
\bibitem[{O'Brien et~al.(2020)O'Brien, Wagner, Schweiker, Mahdavi, Day,
  Kj{\ae}rgaard, Carlucci, Dong, Tahmasebi, Yan, Hong, Gunay, Nagy, Miller, and
  Berger}]{OBrien2020-ns}
\bibinfo{author}{W.~O'Brien}, \bibinfo{author}{A.~Wagner},
  \bibinfo{author}{M.~Schweiker}, \bibinfo{author}{A.~Mahdavi},
  \bibinfo{author}{J.~Day}, \bibinfo{author}{M.~B. Kj{\ae}rgaard},
  \bibinfo{author}{S.~Carlucci}, \bibinfo{author}{B.~Dong},
  \bibinfo{author}{F.~Tahmasebi}, \bibinfo{author}{D.~Yan},
  \bibinfo{author}{T.~Hong}, \bibinfo{author}{H.~B. Gunay},
  \bibinfo{author}{Z.~Nagy}, \bibinfo{author}{C.~Miller},
  \bibinfo{author}{C.~Berger},
\newblock \bibinfo{title}{Introducing {IEA} {EBC} annex 79: Key challenges and
  opportunities in the field of occupant-centric building design and
  operation},
\newblock \bibinfo{journal}{Build. Environ.} \bibinfo{volume}{178}
  (\bibinfo{year}{2020}) \bibinfo{pages}{106738}.
\bibitem[{{Duarte Roa} et~al.(2020){Duarte Roa}, Schiavon, and
  Parkinson}]{targeted-surveyDuarte2020}
\bibinfo{author}{C.~{Duarte Roa}}, \bibinfo{author}{S.~Schiavon},
  \bibinfo{author}{T.~Parkinson},
\newblock \bibinfo{title}{{Targeted occupant surveys : A novel method to
  effectively relate occupant feedback with environmental conditions}},
\newblock \bibinfo{journal}{Preprint}  (\bibinfo{year}{2020}).
\bibitem[{Lipczynska et~al.(2018)Lipczynska, Schiavon, and
  Graham}]{Lipczynska2018}
\bibinfo{author}{A.~Lipczynska}, \bibinfo{author}{S.~Schiavon},
  \bibinfo{author}{L.~T. Graham},
\newblock \bibinfo{title}{{Thermal comfort and self-reported productivity in an
  office with ceiling fans in the tropics}},
\newblock \bibinfo{journal}{Building and Environment} \bibinfo{volume}{135}
  (\bibinfo{year}{2018}) \bibinfo{pages}{202--212}.
\bibitem[{Quintana and Miller(2019)}]{GAN-Quintana2019}
\bibinfo{author}{M.~Quintana}, \bibinfo{author}{C.~Miller},
\newblock \bibinfo{title}{{Poster Abstract: Towards Class-Balancing Human
  Comfort Datasets with GANs}},
\newblock in: \bibinfo{booktitle}{BuildSys '19 Proceedings of the 6th ACM
  International Conference on Systems for Energy-Efficient Built Environments},
  \bibinfo{address}{New York, NY, USA}.
\bibitem[{Abdelrahman et~al.(2020)Abdelrahman, Chong, and
  Miller}]{abdelrahmanbuild2vec}
\bibinfo{author}{M.~M. Abdelrahman}, \bibinfo{author}{A.~Chong},
  \bibinfo{author}{C.~Miller},
\newblock \bibinfo{title}{{Build2Vec: Building Representation in Vector
  Space}},
\newblock in: \bibinfo{booktitle}{SimAUD 2020}, \bibinfo{address}{Online}, pp.
  \bibinfo{pages}{101--104}.

\end{thebibliography}

\end{document}